\documentclass[%
 aip,
 amsmath,amssymb,
 reprint,floatfix%
]{revtex4-1}

\usepackage{color}
\usepackage{graphicx}
\usepackage{dcolumn}
\usepackage{bm}
\usepackage{textcomp}

\usepackage[utf8]{inputenc}
\usepackage[T1]{fontenc}
\usepackage{mathptmx}
\usepackage{etoolbox}
\usepackage{subcaption}
\makeatletter
\def\@email#1#2{%
 \endgroup
 \patchcmd{\titleblock@produce}
 {\frontmatter@RRAPformat}
 {\frontmatter@RRAPformat{\produce@RRAP{*#1\href{mailto:#2}{#2}}}\frontmatter@RRAPformat}
 {}{}
}%
\makeatother
\begin{document}

\title[Quadrature-based Lattice Boltzmann model for non-equilibrium dense gas flows]{Quadrature-based Lattice Boltzmann model for non-equilibrium dense gas flows}
\author{S. Busuioc}
\affiliation{Department of Physics, West University of Timisoara Bd. Vasile Parvan 4, 300223 Timisoara, Romania.}
\affiliation{Institute for Advanced Environmental Research, West University of Timişoara, 300223
Timişoara, Romania}
\email{sergiu.busuioc@e-uvt.ro}

\date{\today}

\begin{abstract}

The Boltzmann equation becomes invalid as the size of gas molecules is comparable with the average intermolecular distance. A better description is provided by the Enskog collision operator, which takes into account the finite size of gas molecules. This extension implies non-local collisions as well as an increase in collision frequency, making it computationally expensive to solve. An approximation of the Enskog collision operator, denoted the simplified Enskog collision operator, is used in this work to develop a quadrature-based Lattice Boltzmann model for non-ideal monatomic dense gases. The Shakhov collision term is implemented in order to fine-tune the Prandtl number. This kinetic model is shown to be able to tackle non-equilibrium flow problems of dense gases, namely the sound wave and the shock wave propagation. The results are compared systematically with the results of the more accurate but computationally intensive particle method of solving the Enskog equation. The model introduced in this paper is shown to have good accuracy for small to moderate denseness of the fluid (defined as the ratio of the molecular diameter to the mean free path) and, due to the efficiency in terms of the computational time, it is suitable for practical applications.

\end{abstract}

\maketitle

\section{\label{sec:intro}Introduction}

Over the past decades, flows at non-negligible values of the Knudsen number Kn (defined as the ratio between the mean free path of the fluid particles in a gas and the characteristic length of the domain), i.e. rarefied gas flows, were successfully approached within the framework of the Boltzmann equation, where the fluid constituents are point particles. The effect of the finite molecular size must be considered when the mean-free path of the fluid particles is comparable to their molecular size\cite{FK72}. This is found in many applications, including high-pressure shock tubes\cite{PH01}, flows through microfabricated nanomembranes\cite{HPWSAGNB06}, single-bubble sonoluminescence\cite{BHL02}, gas extraction in unconventional reservoirs\cite{WLRZ16,SPC17} and the interfacial dynamics of liquid–vapour in high-pressure liquid injection systems\cite{DO15}.

In principle, the Enskog equation can be used to extend the kinetic theory description of fluids to densities beyond the dilute-gas Boltzmann limit\cite{cowling70,FK72,K10}. While keeping binary collision dynamics, the gas molecules are no longer treated as point-like particles, as in the Boltzmann approach, and the finite-size effects are accounted for by including the space correlations between colliding molecules, the molecular mutual shielding and the reduction in the volume available to molecules. This equation can be solved numerically using a probabilistic or deterministic method, just as in the case of the Boltzmann equation. In the past years, the Enskog equation was solved deterministically using different methods, such as the Monte Carlo quadrature method\cite{FS93}('direct method'), the fast spectral method\cite{WZR15,WLRZ16} and the Fokker-Planck approximation\cite{SG17,SG19}. On the other hand, after the success of the Direct Simulation Monte Carlo method (DSMC)\cite{B76}, probabilistic methods have been developed by Alexander et al.\cite{AGA95}, Montanero et al.\cite{MS96} and Frezzotti\cite{F97b} in the $'90$s.
The Enskog equation has been used over the years to study the properties of the hard-sphere dense gas near the solid walls of micro- and nano-channels\cite{D87,DM97,F97b,NFSJMH06}. Its extension to systems of weakly attracting hard-spheres has successfully been used to describe liquid–vapour flows of monoatomic\cite{FGL05,KKW14,FBG19,BGLS20} and polyatomic\cite{Bruno2019,BG20}, mixtures\cite{KSKFW17}, as well as the formation and breakage of liquid menisci in nanochannels\cite{BFG15}.

The methods mentioned above, albeit reliable and accurate, require high computational costs which renders them impractical for many applications. In order to reduce the computational costs, one can simplify the non-local Enskog collision integral by expanding it into a Taylor series around the point $\bm{x}$ in the coordinate space. The first term in this expansion renders the usual Boltzmann collision operator, while the second term is further simplified by replacing the distribution function with the local equilibrium distribution function, which is valid when the fluid is not far from equilibrium\cite{cowling70,K10}. This simplification was used in Lattice Boltzmann (LB) models to investigate non-ideal gases\cite{L98,L00,MM08} and multiphase flows by adding the long-range attractive force\cite{HD02}.
More recently, the simplified Enskog collision operator was successfully implemented in a series of solvers, namely the discrete velocity method\cite{WWHLLZ20}, the discrete unified gas kinetic scheme (DUGKS)\cite{CWWC22arxiv}, the double-distribution LB model\cite{HWA21} and the discrete Boltzmann method\cite{ZXQWW20,GXLLSS22}. They were used to investigate the normal shock wave structures, the rarefaction effects in head-on collisions of two identical droplets and the liquid-vapour phase transition, respectively.


In this paper, we employ a LB model based on Gauss-Hermite quadratures\cite{SYC06}, where finite difference schemes are used for the advection and time-stepping\cite{PBFAS14,AS16a,AS16b,SBBAGL18,ASS20,BABS20}. This finite-difference Lattice Boltzmann (FDLB) belongs to the off-lattice LB models family, which also includes finite-volume and interpolation schemes\cite{X97,C98}. In this approach, the kinetic equation is used to obtain an accurate evolution of the macroscopic moments of $f$\cite{S18}, with less attention directed to the distribution $f$ itself. This allows the momentum space to be optimally sampled for the recovery of the moments of $f$\cite{SYC06}. By using the Gauss quadrature method in the momentum space, off-lattice LB models of any order\cite{SYC06,AS16a,AS16b} can be constructed to accommodate the problem at hand.

This paper is organised as follows. In sec. \ref{sec:ensk_model}, the simplified Enskog equation is presented along the FDLB model used to numerically solve it. The particle method of solving the Enskog equation\cite{F97b}, which is used to systematically compare the FDLB results in the case of the shock wave propagation, is briefly presented in Sec.~\ref{sec:ensk_dsmc}. The simulation results are reported in Sec.~\ref{sec:results}.
In Sec.~\ref{sec:results_sound_wave} the sound wave propagation results are compared with the analytic solution, while in Sec.~\ref{sec:results_shock_wave}, the shock wave results are compared with the results obtained using the particle method, as well as with the inviscid limit solution. We conclude the paper in Sec.\ref{sec:conclusions}. The details regarding the numerical schemes employed in this paper, namely the third-order TVD Runge-Kutta method for time-stepping, the fifth-order WENO-5 advection scheme and the $6$th order central difference scheme used for gradient evaluation, are relegated to Appendix~\ref{app:numerical_scheme}

\section{\label{sec:ensk_model}The Enskog Lattice Boltzmann model}

\subsection{\label{sec:ensk_model_equation}Enskog equation}

The Enskog equation describing the evolution of a system composed of rigid spherical molecules was proposed by its author in 1922\cite{enskog22}. Unlike Boltzmann in his equation, where molecules are assumed to be point-like particles and collisions are local, Enskog has taken into account the volume of the fluid particles (i.e., molecules,) that reduces the free movement space available to each particle, which results into an increased number of collisions. Moreover, the interparticle collisions are non-local, as the positions of the two colliding molecules are one molecular diameter apart. The Enskog equation can be written as\cite{cowling70,K10}:
\begin{equation}\label{eq:enskog}
 \frac{\partial f}{\partial t}+\frac{\bm{p}}{m} \cdot\bm{\nabla}_{\bm{x}} f + \bm{F}\cdot\bm{\nabla}_{\bm{p}} f =J_E
\end{equation}
where $m$ is the particle mass, ${\bm{F}}$ is the external body force and $f(\bm{x},\bm{p},t)$ is the single-particle distribution function, giving at time $t$ the number of particles of momentum $\bm{p}$ located within the unit phase space volume centered in the point whose position vector is $\bm{x}$.
The right-hand side is given by the Enskog collision operator $J_E$ which reads:
\begin{multline}\label{eq:collision_integral}
 J_E=\sigma^2\int \left\{ \chi\left({\bm{x}}+\frac{\sigma}{2}{\bm{k}}\right)f({\bm{x}},\bm{p^*})f({\bm{x}} + \sigma {\bm{k}},\bm{p_1^*}) \right.\\ -
        \left.      \chi\left({\bm{x}}-\frac{\sigma}{2}{\bm{k}}\right)f({\bm{x}},\bm{p})f({\bm{x}} - \sigma {\bm{k}},\bm{p_1}) \right\}
        ({\bm{p_r}}\cdot{\bm{k}}) d{\bm{k}}d{\bm{p_1}}
\end{multline}
where $\sigma$ is the molecular diameter. $\bm{p_r}=\bm{p_1}-\bm{p}$ is the relative momentum and ${\bm k}$ is the unit vector giving the relative position of the two colliding particles. In the equation above, the distribution function dependence on time $t$ was dropped for brevity. The superscript $*$ refers to the post-collision momenta.

The contact value of the pair correlation function $\chi$ accounts for the effect of the molecular diameter $\sigma$ on the collision frequency. In the standard Enskog theory (SET), $\chi$ is approximated by the value of the pair correlation function at the contact point of two colliding particles in a fluid which is in uniform equilibrium. An approximate, but accurate expression for $\chi_{\mbox{\tiny SET}}$, namely:
\begin{equation}\label{eq:chi}
 \chi_{\text{\tiny SET}}[n]=\frac{1}{nb}\left(\frac{P^{hs}}{n k_B T}-1\right)=\frac{1}{2}\frac{2-\eta}{(1-\eta)^3},
\end{equation}
is obtained from the equation of state of the hard-sphere fluid proposed by Carnahan and Starling~\cite{CS69}:
\begin{equation}
 P^{hs}=nk_B T\frac{1+\eta+\eta^2-\eta^3}{(1-\eta)^3}
\end{equation}
where $n$ is the particle number density, $\eta=b \rho /4$ is the reduced particle density, with $b=2\pi\sigma^3/3m$, $p^{hs}$ is the pressure of a system of hard-spheres and $k_B$ is the Boltzmann constant and $T$ is the temperature. The square brackets in Eq.~\eqref{eq:chi} denote a functional dependence.

In the revised (modified) Enskog theory\cite{vBE73}, $\chi$ is given by the value of the pair correlation function at the contact point of the two colliding particles in a fluid in non-uniform equilibrium. A good approximation for the radial distribution function is obtained following the Fischer-Methfessel (FM) prescription~\cite{FM80}. In this approach, the actual value of the density at the contact point is replaced with $\overline{n}({\bm{x}})$, which represents the value of the density field averaged over a spherical volume of radius $\sigma$ centered in the point ${\bm x}$. Consequently, the contact value of the pair correlation function is given by:
\begin{subequations}
\begin{equation}\label{eq:chi_ret}
\chi_{\mbox{\tiny RET-FM}}\left(n\Big(\bm{x}-\frac{\sigma}{2} {\bm{\hat{k}}}\Big)\right)=\chi_{\mbox{\tiny SET}}\left(\overline{n}\Big(\bm{x}-\frac{\sigma}{2} {\bm{\hat{k}}}\Big)\right),
\end{equation}
where
\begin{eqnarray}
\overline{n}(\bm{x})&=&\frac{3}{4\pi \sigma^3}\int_{\mathbb{R}^3} n(\bm{x}')w(\bm{x},\bm{x}')\,d\bm{x}', \\
w(\bm{x},\bm{x}')&=&\left\{
\begin{array}{cc} 
1, &\qquad \|\bm{x}'-\bm{x}\|<\sigma, \\
0, & \qquad \|\bm{x}'-\bm{x}\|>\sigma.
\end{array}
\right.
\end{eqnarray}
\end{subequations}

The Enskog collision operator in Eq.~\eqref{eq:collision_integral} can be regarded as a generalisation of the Boltzmann collision operator to particles that have spatial extent. By taking the limit of molecular diameter $\sigma$ going to zero, the pair correlation function goes to unity ($\chi\rightarrow1$) and one obtains the Boltzmann collision operator since the term $\sigma^2$ stems from the scattering cross-section.

We base the non-dimensionalization procedure employed in this paper on reference quantities\cite{AS18}, which we introduce as follows. Let $L_{\text{ref}}$ be the value of the reference length. The reference values of the particle number density and the temperature are denoted $n_{\text{ref}}$ and $T_{\text{ref}}$, respectively. Hence the reference value of the momentum is $p_{\text{ref}}=\sqrt{m_{\text{ref}} k_B T_{\text{ref}}}$ and the reference time is $t_{\text{ref}} = m_{\text{ref}} L_{\text{ref}}/p_{\text{ref}}$, where $m_{\text{ref}}$ is the mass of a fluid particle.

\subsection{\label{sec:ensk_model_shakhov}Enskog-Shakhov equation using the simplified Enskog collision operator}

By assuming that the contact value of the pair correlation function $\chi$ (functional dependence dropped for brevity) and the distribution functions $\{f^*\equiv f({\bm{x}},\bm{p^*}),f_1^*\equiv f({\bm{x}} + \sigma {\bm{k}},\bm{p_1^*}),f\equiv f({\bm{x}} ,\bm{p}) ,f_1\equiv f({\bm{x}} - \sigma {\bm{k}},\bm{p_1})\}$ are smooth functions, one can approximate these functions in the Enskog collision integral $J_E$ through a Taylor series near the point $\bm{x}$. The resulting terms up to first order gradients $J_E\approx J_0+J_1$ are\cite{cowling70,K10}:
\begin{eqnarray}
 J_0(f,f) &=& \chi\int (f^*f_1^*-ff_1)\sigma^2({\bm{p_r}}\cdot{\bm{k}}) d{\bm{k}}d{\bm{p_1}}\\
 J_1(f,f) &=& \chi\sigma\int\bm{k}(f^*\bm{\nabla}f_1^*-f\bm{\nabla} f_1)\sigma^2({\bm{p_r}}\cdot{\bm{k}}) d{\bm{k}}d{\bm{p_1}}\nonumber\\
 &+&\frac{\sigma}{2}\int\bm{k}\bm{\nabla}\chi(f^*f_1^*-ff_1)\sigma^2({\bm{p_r}}\cdot{\bm{k}}) d{\bm{k}}d{\bm{p_1}}
\end{eqnarray}
where all functions $f^*,f_1^*,f,f_1$ and $\chi$ are evaluated at the point ${\bm{x}}$.

The collision term $J_0(f,f)$ is the usual collision term of the Boltzmann equation multiplied by $\chi$, and is treated as such, by applying the usual relaxation time approximation. In this paper we will employ the Shakhov collision term\cite{shakhov68a,shakhov68b}, namely:
\begin{equation}
 J_0(f,f)=-\frac{1}{\tau}(f-f^S),
\end{equation}
where $\tau$ is the relaxation time and $f_S$ is the equilibrium Maxwell-Boltzmann distribution times a correction factor\cite{shakhov68a,shakhov68b,GP09,ASS20}:
\begin{equation}
 f^S=f_{\text{\tiny MB}}\left[1 + \frac{1-\text{Pr}}{P_i k_B T}\left( \frac{\bm{\xi}^2}{5 m k_B T}-1 \right)\bm{\xi}\cdot \bm{q} \right]
\end{equation}
where $q$ is the heat flux obtained using:
\begin{equation}
 \bm{q}=\int d^3p f \frac{\bm{\xi}^2}{2m}\frac{\bm{\xi}}{m},
\end{equation}
$\bm{\xi}=\bm{p}-m\bm{u}$ is the peculiar momentum, $\text{Pr}=c_P\mu/\lambda$ is the Prandtl number, $c_P=5k_B/2m$ is the specific heat at constant pressure and $P_i=\rho R T=n k_B T$ is the ideal gas equation of state, with $R$ being the specific gas constant. The Maxwell-Boltzmann distribution $f_{\text{\tiny MB}}$ is given by:
\begin{equation}
 f_{\text{\tiny MB}}=\frac{n}{(2 m \pi k_B T)^{3/2}}\exp{\left(-\frac{\bm{\xi}^2}{2 m k_B T}\right)}
\end{equation}

The second term of $J_E$, namely $J_1(f,f)$, can be approximated by replacing the distribution functions ($f^*,f_1^*,f,f_1$) with the corresponding equilibrium distribution functions. By using $f_{\text{\tiny MB}}^*f_{\text{\tiny MB},1}^*=f_{\text{\tiny MB}}f_{\text{\tiny MB},1}$, and integrating over $\bm{k}$ and $\bm{p_1}$, one obtains\cite{cowling70,K10}:
\begin{multline}
 J_1(f,f)\approx J_1(f_{\text{\tiny MB}},f_{\text{\tiny MB}})=\\-b \rho \chi f_{\text{\tiny MB}} \left\{\bm{\xi}\left[\bm{\nabla}\ln(\rho^2 \chi T)+\frac{3}{5}\left(\zeta^2-\frac{5}{2}\right)
 \bm{\nabla}\ln T\right]\right.\\
 \left. + \frac{2}{5}\left[ 2\bm{\zeta}\bm{\zeta}\bm{:\nabla u} + \left(\zeta^2-\frac{5}{2} \right)\bm{\nabla\cdot u} \right]
 \right\}
\end{multline}
where $\bm{\zeta}=\bm{\xi}/\sqrt{2RT}$. With the above approximations and considering no external force, the Enskog equation Eq.~\eqref{eq:enskog} becomes:
\begin{equation}\label{eq:enskog_approx}
 \frac{\partial f}{\partial t}+\frac{\bm{p}}{m}\nabla_{\bm{x}}f=-\frac{1}{\tau}(f-f_S)+J_1(f_{\text{\tiny MB}},f_{\text{\tiny MB}})
\end{equation}

The macroscopic quantities are evaluated as moments of the distribution function:
\begingroup
\renewcommand*{\arraystretch}{1.25}

\begin{equation}
 \begin{pmatrix}
 n \\ \rho\bm{u} \\ \frac{3}{2} n k_B T
 \end{pmatrix} =\int d^3 p
 \begin{pmatrix}
 1 \\ \bm{p} \\ \frac{\bm{\xi}^2}{2m}
 \end{pmatrix} f
\end{equation}
\endgroup
where $\rho=mn$.

The Chapman-Enskog expansion of Eq.~\eqref{eq:enskog_approx} yields the following conservation equations for mass, momentum and energy\cite{K10}:
\begin{subequations}
\begin{align}
 \frac{D\rho}{Dt}+\rho\nabla \bm{u}&=0\\
 \rho \frac{D {\bm{u}}}{Dt}+\nabla P&=-\nabla\cdot\Pi \\
 \rho\frac{De}{Dt}+P\nabla\cdot\bm{u}&=-\nabla\cdot \bm{q}+\Pi\bm{:}\nabla\bm{u}
 \end{align}
\end{subequations}
where $D/Dt=\partial_t+\bm{u}\cdot\nabla$ is the material derivative and
$P=P_i(1+b\rho\chi)$ is the equation of state of a non-ideal gas. The heat flux and
the viscous part of the stress tensor $\Pi_{\alpha\beta}$ are given by:
\begin{eqnarray}
&\bm{q}=-\lambda\nabla T, \\
&\hspace{-10pt}\Pi=-\mu_v\mathcal{I}\bm{\nabla} \cdot \bm{u}-\mu\left(\nabla u + (\nabla u)^T-\frac{2}{3}\mathcal{I}\bm{\nabla} \cdot \bm{u}\right)
\end{eqnarray}
where $\mathcal{I}$ is the identity matrix and the bulk viscosity $\mu_v$, the shear viscosity $\mu$ and the thermal conductivity $\lambda$ are given by\cite{K10}:
\begin{subequations}
\begin{equation}
\mu_v=\frac{16}{5\pi}\mu_0 b^2 \rho^2 \chi\label{eq:bulk_viscosity},
\end{equation}
\begin{equation}
\mu=\tau P_i=\mu_0 b\rho\left[\frac{1}{b\rho\chi}+0.8+\frac{4}{25}\left(1+\frac{12}{\pi}\right)b\rho\chi\right]\label{eq:viscosity},
\end{equation}
\begin{equation}
 \lambda=\frac{5k_B}{2m}\frac{\tau P_i}{\text{Pr}}=\lambda_0 b\rho\left[\frac{1}{b\rho\chi}+1.2+\frac{9}{25}\left(1+\frac{32}{9\pi}\right)b\rho\chi\right],
\end{equation}
\end{subequations}
In these equations, $\mu_0=\mu_{\text{ref}}\sqrt{T/T_0}$ is the viscosity coefficient for hard-sphere molecules, where $\mu_{\text{ref}}$ represents the viscosity coefficient for dilute gases at temperature $T_0$ and $\lambda_0\equiv\lambda_{\text{ref}}$ is the reference thermal conductivity for dilute gases at temperature $T_0$. The reference values are\cite{K10}:
\begin{equation}
 \mu_{\text{ref}}=\frac{5}{16\sigma^2}\sqrt{\frac{m k_B T_0}{\pi}},\quad \lambda_{\text{ref}}=\frac{75 k_B}{64 m\sigma^2}\sqrt{\frac{m k_B T_0}{\pi}}.
\end{equation}

For the dense gas the Prandtl number is:
\begin{equation}
 \text{Pr}=\frac{2}{3}\,\frac{1+\frac{4}{5}b\rho\chi+\frac{4}{25}\left(1+\frac{12}{\pi}\right)(b\rho\chi)^2}{1+\frac{6}{5}b\rho\chi+\frac{9}{25}\left(1+\frac{32}{9\pi}\right)(b\rho\chi)^2}
\end{equation}
with the dilute limit of $\text{Pr}=2/3$.

From here it follows directly that the relaxation time $\tau$ is given by:
\begin{equation}
 \tau=\frac{\mu}{P_i}
\end{equation}

Since $\mu$ takes into account both the kinetic and the potential contributions, associated with the flow of the molecules and collisional contribution to the transfer of gas momentum and energy\cite{cowling70,K10}, respectively, the collisional transfer due to the non-local molecular collisions is well described in the relaxation time approximation. Note that the viscosity of the dense gas of a fixed reduced density $\eta$ can be changed by varying the molecular diameter $\sigma$ and the number density $n$.

By using the reference mean free path $l=m/\sqrt{2}\pi\sigma^2 n \chi$, one can define the degree of denseness $E_l$ introduced by Frezzotti and Sgarra\cite{FS93}, given by the ratio of the molecular diameter and the mean free path:
\begin{equation}
 E_l=\frac{\sigma}{l}=\frac{3}{\sqrt{2}}b n \chi.
\end{equation}

The relaxation time $\tau$ can be rewritten as the molecular diameter $\sigma$ times a functional $g$ of $\eta$:
\begin{equation}\label{eq:tau_sigma}
 \tau=\sigma g[\eta]
\end{equation}
such that one can vary $\tau$ at constant reduced density $\eta$ by changing $\sigma$. Furthermore, in the case of the standard Enskog theory (SET), one can keep $\sigma$ and $n$ fixed (i.e. a constant $\eta$) and multiply $\tau$ with a relaxation scaling factor $\widetilde{\tau}$, which is equivalent to setting $\sigma=\widetilde{\tau}$ and keeping $\eta$ constant. This is true also for $J_1$ since all terms remain unchanged when $\eta=\text{const}$ and varying $\sigma$ and $n$.

\subsection{Reduced distributions}\label{sec:reduced_distribution}

In the context of the longitudinal waves and 1D shock waves considered in this paper, the dynamics along the $y$ and $z$ directions is trivial and it is convenient to integrate out the momentum space degrees of freedom at the level of the model equation.
The $y$ and $z$ degrees of freedom can be integrated out and two reduced distribution functions, $\phi$ and $\theta$, can be introduced as\cite{LZ04,GP09,MWRZ13,AS18,AS2019,BA19}:
\begin{align}
 \phi({\bm x},p_x,t)&=\int dp_y dp_z f({\bm x},{\bm p},t),\\
 \theta({\bm x},p_x,t)&= \int dp_y dp_z \frac{p_y^2+p_z^2}{m}f({\bm x},{\bm p},t)
\end{align}
In the following, all dependencies of the reduced distribution functions will be dropped for brevity. The macroscopic moments can be evaluated as:
\begingroup
\renewcommand*{\arraystretch}{1.2}
\begin{align}
 \begin{pmatrix}
 n \\ \rho u_x \\ \Pi_{xx}
 \end{pmatrix} &=\int d p_x
 \begin{pmatrix}
 1 \\ p_x \\ \frac{\xi_x^2}{m}
 \end{pmatrix} \phi,\\
  \begin{pmatrix}
 \frac{3}{2}n k_B T \\ q_x
 \end{pmatrix} &=\int d p_x
 \begin{pmatrix}
 1 \\ \frac{\xi_x}{m}
 \end{pmatrix} \left(\frac{\xi_x^2}{2m}\phi+\frac{1}{2}\theta\right)
\end{align}
\endgroup

The evolution equations for the reduced distribution functions are:
\begingroup
\renewcommand*{\arraystretch}{1.25}
\begin{equation}\label{eq:evolution}
 \frac{\partial}{\partial t}
 \begin{pmatrix}
 \phi\\ \theta
 \end{pmatrix}
+ \frac{p_x}{m}\frac{\partial}{\partial x}
\begin{pmatrix}
 \phi\\ \theta
 \end{pmatrix}
= -\frac{1}{\tau} \begin{pmatrix}
 \phi-\phi_S\\ \theta-\theta_S
 \end{pmatrix}
 + \begin{pmatrix}
 J_1^\phi\\ J_1^\theta
 \end{pmatrix}
 \end{equation}
\endgroup

In the above the, $\phi_S$ and $\theta_S$ are given by:
\begin{align}
\phi_S=f^x_{\text{\tiny MB}}\left[1 + \frac{1-\text{Pr}}{5P_i m k_B T}\left( \frac{\xi_x^2}{m k_B T}-3 \right)\xi_x q_x \right],\\
\theta_S = 2 k_B T f^x_{\text{\tiny MB}}\left[1 + \frac{1-\text{Pr}}{5P_i m k_B T}\left( \frac{\xi_x^2}{m k_BT}-1 \right)\xi_x q_x \right]
\end{align}
where
\begin{equation}\label{eq:f_mb}
 f^x_{\text{\tiny MB}}=\frac{n}{(2 m \pi k_B T)^{1/2}}\exp{\left(-\frac{\xi_x^2}{2 m k_B T}\right)}
\end{equation}
while the first order corrections $J_1^\phi$ and $J_1^\theta$ are:
\begin{subequations}\label{eq:j1_expanded}
 \begin{multline}
 J_1^\phi=-\left[ \xi_x\partial_x\ln\chi + 2\xi_x\partial_x\ln\rho + \frac{3}{5}\left(\frac{\xi_x^2}{m k_B T}-1 \right)\partial_x u_x \right. \\ \left.+\frac{3}{10}\left(\frac{\xi_x^3}{m^2 k_B T}+\frac{\xi_x}{3m} \right)\partial_x\ln T
 \right]f_{\text{\tiny MB}}
 \end{multline}
\begin{multline}
  J_1^\theta=-\left[ \xi_x\partial_x\ln\chi + 2\xi_x\partial_x\ln\rho + \frac{3}{5}\left(\frac{\xi_x^2}{m K_B T}-\frac{1}{3} \right)\partial_x u_x \right. \\ \left.+\frac{3}{10}\left(\frac{\xi_x^3}{m^2 k_B T}+\frac{7\xi_x}{3m} \right)\partial_x\ln T
 \right]2 m k_B Tf_{\text{\tiny MB}}b\rho\chi
\end{multline}
\end{subequations}

\subsection{\label{sec:ensk_model_lb}The finite-difference Enskog Lattice Boltzmann model}

By using the reduced distribution, one has to solve the 1D evolution equations Eqs.~\eqref{eq:evolution}. In the following, we will introduce the notation $\psi\in\{\phi,\,\theta\}$ to represent the reduced distributions introduced in Sec.~\ref{sec:reduced_distribution}.


When the Shakhov collision term is used in an LB model, the moments of the distribution function $\psi({x},{p},t)$ up to order $N \geq 6$ are needed in order to get the evolution equations of the macroscopic fields\cite{AS18}.
Thus, the minimum number of the momentum vectors in the LB model based on the full-range Gauss-Hermite quadrature that ensures all the moments of $\psi({x},{p_x},t)$ up to order $N_{\text{min}}=6$ is $Q_{\text{min}}=(N_{\text{min}}+1)=7$\cite{SYC06,PBFAS14,FSFBSLA15,AS16a}. Hence, the momentum set $\{{p}_k\}$ has $Q \geq Q_{\text{min}}$ elements that belong to the set $\{{r}_{k}\}$, $1\leq k \leq Q$, of the roots of the full-range Hermite polynomial $H_{Q}(p)$\cite{SYC06,AS16a} and the their associated weights ${w}_{k}$ given by\cite{AS16a,AS16b,H87,OLBC10}
\begin{equation}
{w}_{k} = \frac{Q !}{\,[H_{Q+1}({r}_{k})]^{2}\,}.
\end{equation}

The full range Hermite polynomials $H_{\ell}(p)$ used in this paper are the so-called {\emph{probabilistic} Hermite polynomials, which are orthogonal with respect to the weight function
\begin{equation}
\omega(p) = \frac{1}{\,\sqrt{2\pi}\,} e^{-p^{2}/2},
\label{eweight}
\end{equation}
and their orthogonality relation reads\cite{H87}
\begin{equation}
 \int_{-\infty}^{+\infty} dp \,\omega(p) H_{\ell}(p) H_{\ell'}(p) =
 \ell ! \, \delta_{\ell,\ell'}.
\end{equation}

The equilibrium functions $f_{\text{\tiny MB}}^{k}\equiv f_{\text{\tiny MB}}(x,p_k,t)$ are replaced by\cite{AS16a,AS16b}:
\begin{subequations}\label{eq:feq}
\begin{equation}
f_{\text{\tiny MB}}^{k} = n g_{k},
\end{equation}
where
\begin{equation}
 g_{k} \equiv g_{k}\left[ u,
 T \right] =
 w_{k}\,\sum_{\ell=0}^{N}
 H_{\ell}(p_{k}) \sum_{s=0}^{\lfloor \ell/2 \rfloor}
\frac{ \,( mT -1 )^{s} ( mu )^{\ell-2s}\,}{\,2^{s} s! (\ell-2s)!\,}
\label{g_alpha},
\end{equation}
\end{subequations}
and $\lfloor \ell/2 \rfloor$ is the integer part of $\ell/2$.

The non-dimensionalized form of the evolution equation of the functions $\phi_k$ and $\theta_k$ is:
\begingroup
\renewcommand*{\arraystretch}{1.25}
\begin{equation}\label{eq:evolution_lb}
 \frac{\partial}{\partial t}
 \begin{pmatrix}
 \phi_k\\ \theta_k
 \end{pmatrix}
+ \frac{p_k}{m}\frac{\partial}{\partial x}
\begin{pmatrix}
 \phi_k \\ \theta_k
 \end{pmatrix}
= -\frac{1}{\tau} \begin{pmatrix}
 \phi_k-\phi_{S;k}\\ \theta_k-\theta_{S;k}
 \end{pmatrix}
 + \begin{pmatrix}
 J_{1;k}^\phi\\ J_{1;k}^\theta
 \end{pmatrix}.
 \end{equation}
\endgroup

The macroscopic quantities are evaluated as:
\begingroup
\renewcommand*{\arraystretch}{1.25}
\begin{align}
 \begin{pmatrix}
 n \\ \rho u \\ \Pi
 \end{pmatrix} &=\sum_{k=1}^Q
 \begin{pmatrix}
 1 \\ p_k \\ \frac{\xi_k^2}{m}
 \end{pmatrix} \phi_k,\\
  \begin{pmatrix}
 \frac{3}{2} n k_B T \\ q
 \end{pmatrix} &=\sum_{k=1}^Q d p_k
 \begin{pmatrix}
 1 \\ \frac{\xi_k}{m}
 \end{pmatrix} \left(\frac{\xi_k^2}{2m}\phi_k+\frac{1}{2}\theta_k\right)
\end{align}
\endgroup

\section{\label{sec:ensk_dsmc} Particle method for Enskog equation}

The Enskog equation Eq.~\eqref{eq:enskog} is solved numerically using also a particle method. The method is an extension of the original Direct Simulation Monte-Carlo (DSMC) to deal with the nonlocal structure of the Enskog collision integral\cite{F97b}. For a thorough description of the numerical scheme and the analysis of its computational complexity please refer to Ref.~\cite{FBG19}. A brief description of the scheme is outlined below.

The main framework of the DSMC scheme used to solve the Boltzmann equation is preserved, with modifications occurring in the collision algorithm due to the nonlocal structure of the Enskog collision operator.
The distribution function is represented by $N$ computational particles:
\begin{equation}
f(\bm{x},\bm{p},t)= \frac{1}{m} \sum_{i=1}^{N} \delta{\left(\bm{x}-\bm{x}_i(t)\right)} \delta(\bm{p}-\bm{p}_i(t)),
\end{equation}
where $\bm{x}_i$ and $\bm{p}_i$ are the position and the momentum of the $i$th particle at time $t$, respectively.

The distribution function is updated by a fractional-step method based on the time-splitting of the evolution operator in two sub-steps, namely free streaming and collision. In the first stage, the distribution function is advanced from $t$ to $t+\Delta t$ by neglecting the collisions between particles, i.e. by solving the equation:
\begin{equation}\label{eq:stage_I}
  \frac{\partial f}{\partial t} +\frac{\bm{p}}{m}\cdot\nabla_{\bm{x}}f=0,
\end{equation}
which translates into updating the positions of the computational particles according to:
\begin{equation}
 \bm{x}_i(t+\Delta t)=\bm{x}_i(t)+\frac{\bm{p}_i}{m}\Delta t,
\end{equation}
with the resulting distribution function denoted $\tilde{f}(\bm{x},\bm{p},t+\Delta t)$.

In the second stage, the short-range hard-sphere interactions are evaluated and the updating rule for the distribution function is given by:
\begin{equation}
 f(\bm{x},\bm{p},t+\Delta t)= \tilde{f}(\bm{x},\bm{p},t+\Delta t)+J_E[\tilde{f}]\Delta t.
\end{equation}
During this stage, the $N$ particle positions $\bm{x}_i$ are unchanged while their momenta $\bm{p}_i/m$ are modified according to stochastic rules which essentially correspond to the Monte Carlo evaluation of the collision integral given by Eq.~\eqref{eq:collision_integral} by selecting collision pairs accordingly.
The macroscopic quantities are obtained by time-averaging the particles' microscopic states, as well as phase averaging, by running identically macroscopic but statistically independent simulations (i.e. same initialisation but a different random seed).

\section{\label{sec:results}Results}

\subsection{\label{sec:results_sound_wave}Longitudinal waves}

\begin{figure}
  \includegraphics[width=\linewidth]{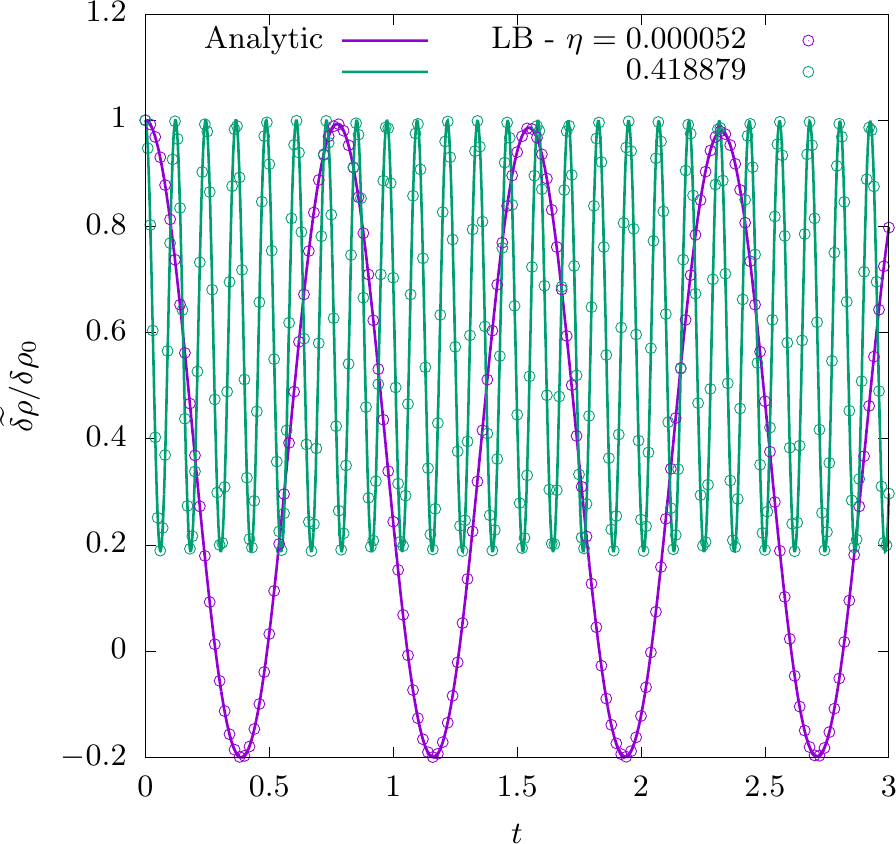}
  \caption{The evolution of the normalized density amplitude $\delta\rho(t)/\delta\rho_0$ obtained numerically with $N_x=100$ and $Q_x=8$, compared with the analytic prediction in Eq.~\eqref{eq:sw_solution}, for two values of the reduced density $\eta$.}
  \label{fig:sound_wave_time_evolution}
\end{figure}

\subsubsection{Problem statement}

The study of longitudinal waves is an important topic in fluid mechanics\cite{F95,SK08,WX12,S15,A18,S19}. The propagation of longitudinal waves induces fluctuations in the macroscopic properties of the fluid, the amplitudes of which decay due to viscous and thermal dissipation. The sound wave propagates as a longitudinal wave through the compression and relaxation of the neighboring fluid elements. For simplicity, we will consider small perturbations of density and pressure around the constant values $\rho_0$ and $P_0$ in a fluid that is homogeneous along the $y$ and $z$ axis. The wave propagates along the $x$ axis with a small velocity $u(x,t)$:
\begin{equation}
 \rho(x,t)=\rho_0[1+\delta\rho(x,t)],\quad P(x,t)=P_0[1+\delta P(x,t)]
\end{equation}
where the perturbations $\delta\rho$ and $\delta P$ are of the same order of magnitude as $u$.

\subsubsection{Analytic solution}

We will briefly go through the usual approach to sound wave propagation\cite{F95,KCD15,W16}.
In the linearised regime, the macroscopic equations reduce to:
\begin{subequations}
\label{eq:sw_linearised}
\begin{align}
 \partial_t\delta \rho +\partial_x u = 0& \label{eq:sw_linearised_density}\\
 \partial_t u + \frac{P_0}{\rho_0}\partial_x \delta P - \frac{1}{\rho_0}\partial_x \Pi = 0& \label{eq:sw_linearised_momentum}\\
 \partial_t \delta T + \frac{\partial_x q}{\rho_0 c_V T_0} + \frac{P_0}{\rho_0 c_V T_0}\partial_x u = 0&\label{eq:sw_linearised_energy}
\end{align}
\end{subequations}
where the specific energy is $e=c_V T=c_V T_0 (1+\delta T)$ and $\Pi=O(u)$.

Considering that the pressure $P$ depends on $x$ and $t$ only through the variables $\rho$ and $T$, the derivative can be written as:
\begin{equation}
P_0\partial_x\delta P = \rho_0(\partial_\rho P )(\partial_x \delta\rho) + T_0(\partial_T P) (\partial_x\delta T)
\end{equation}

By replacing the above results in Eq.~\eqref{eq:sw_linearised_momentum} and applying a time derivative of the whole equation one obtains:
\begin{equation}\label{eq:sw_linearised_energy_II}
 \partial_t^2 u - \left(\partial_\rho + \frac{P_0}{\rho_0^2c_V}\partial_T P \right)\partial_x^2 u=\frac{1}{\rho_0^2 c_V}(\partial_T P)(\partial_x^2 q) + \frac{1}{\rho_0}\partial_t\partial_x \Pi
\end{equation}

By neglecting dissipative effects one can identify the square of the sound speed as:
\begin{equation}\label{eq:sw_cs}
 c_s^2=\partial_\rho P+\frac{P_0}{\rho_0^2 c_V}\partial_T P
\end{equation}

A harmonic decomposition can be performed with respect to the perturbation amplitudes based on the linearity and homogeneity of Eqs.~\eqref{eq:sw_linearised}. Given a wave number $k=2\pi/L$ of a longitudinal wave of length $L$, the following relations can be established:
\begingroup
\renewcommand*{\arraystretch}{1.25}
\begin{equation}
 \begin{pmatrix}
 \delta\rho \\ \delta P \\ \tau
 \end{pmatrix} =
 \begin{pmatrix}
 \widetilde{\delta\rho}(t) \\ \widetilde{\delta P}(t) \\ \widetilde{\tau}(t)
 \end{pmatrix} \cos(k x),\quad
  \begin{pmatrix}
 u \\ q
 \end{pmatrix} =
 \begin{pmatrix}
 \widetilde{u}(t) \\ \widetilde{q}(t)
 \end{pmatrix} \sin(k x)
\end{equation}
\endgroup
where the amplitudes $\widetilde{A}\equiv \widetilde{A}(t)\,(\widetilde{A}\in\{\widetilde{\delta\rho},\widetilde{\delta P},\widetilde{\tau},u,q\})$ depend only on time. These amplitudes can be written in terms of independent modes:
\begin{equation}
 \widetilde{A}(t)=\sum_\alpha e^{-\alpha t}A_\alpha
\end{equation}
where $A_\alpha$ are constants. The viscous part of the stress tensor $\Pi_{ij}$ can be written as:
\begin{equation}
 \Pi=\left( \frac{4\mu}{3}+\mu_V\right)\partial_x u \implies \Pi_\alpha=\left( \frac{4\mu}{3}+\mu_V\right)k u_\alpha
\end{equation}

From the energy equation \eqref{eq:sw_linearised_energy} one gets:
\begin{equation}
 \alpha\delta T_\alpha=\frac{k}{\rho_0 c_V T_0}(q_\alpha+P_0u_\alpha)
\end{equation}

By virtue of the Fourier law $\bm{q}=-\lambda\nabla T$, one obtains:
\begin{equation}
 q_\alpha = \frac{\lambda k^2 P_0}{\alpha\rho c_V -\lambda K^2}u_\alpha.
\end{equation}

Replacing all into Eq.~\eqref{eq:sw_linearised_energy_II} we get:
\begin{multline}
 \alpha^3 - \frac{\mu}{\rho_0}k^2\alpha^2\left( \frac{4}{3} +\frac{\mu_V}{\mu} + \frac{\gamma}{\text{Pr}} \right) + \\
 k^2 c_s^2\alpha \left[1 + \frac{\gamma k^2\mu^2}{\rho_0^2c_s^2 \text{Pr}} \left( \frac{4}{3}+\frac{\mu_V}{\mu}\right) \right]-\frac{\gamma\mu k^4}{\rho_0\text{Pr}}\partial_\rho P=0
\end{multline}
where $\text{Pr}=c_P\mu/\lambda$ and $\gamma=c_s^2\rho/P$ is the adiabatic index.

The above equation is cubic with respect to $\alpha$, thus it admits at least one real solution, which corresponds to the thermal mode $\alpha_t$. The other two roots $\alpha_{\pm}$, corresponding to the acoustic modes, must be complex in order to allow the wave to propagate. Writing $\alpha_{\pm} = \alpha_a \pm i\alpha_s$, we see that $\alpha_a$ induces acoustic dissipation, while $\alpha_s = k c_s$ is related to the speed of sound $c_s$ at the background parameters:
\begin{align}\label{eq:sw_coeff}
 &\alpha_t=\frac{\gamma\mu k^2}{\text{Pr}\rho_0 c_s^2}\partial_\rho P,\quad \alpha_s=k c_s\nonumber\\
 &\alpha_a=\frac{k^2\mu}{2\rho_0}\left[ \frac{4}{3} +\frac{\mu_V}{\mu} + \frac{\gamma c_s^2}{\text{Pr}}\left(1 - \partial_\rho P\right) \right]
\end{align}

In this paper, we will restrict our simulations to the case when the pressure perturbation vanishes at initial time $\delta P(t_0)=0$, since all other combinations are equivalent.
After some calculations one can write the full solution of the density amplitude:
\begin{equation}\label{eq:sw_solution}
 \delta\rho(t) \approx \delta\rho_0\left[ e^{\alpha_t t}+\left(e^{-\alpha_a t}\cos(k c_s t)-e^{-\alpha_t t} \right)\frac{\partial_\rho P}{c_s^2} \right]
\end{equation}

\subsubsection{Computational setup}

\begin{figure}
  \includegraphics[width=\linewidth]{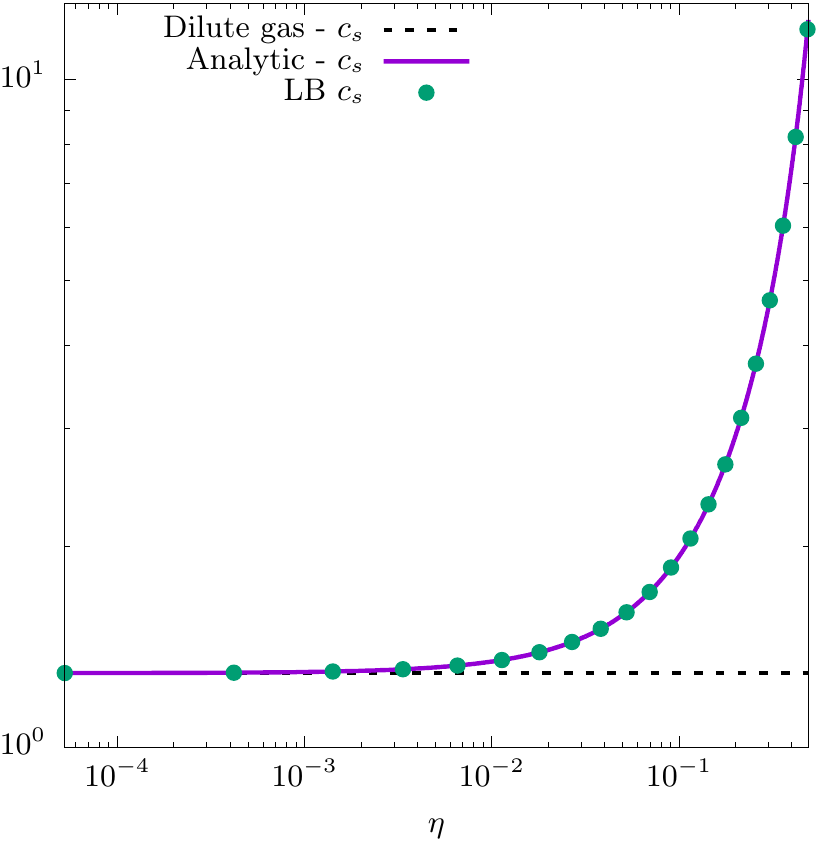}
  \caption{The sound speed $c_s$ obtained from the simulation results and the analytic prediction Eq.~\eqref{eq:sw_cs}.}
  \label{fig:sound_wave_cs}
\end{figure}

All simulations are performed using a system of length equal to $L=1$ ($k=2\pi/L$) and $N_x=100$ nodes. The quadrature order is chosen to be $Q=8$ as it has resulted to be more stable than the minimal $Q=7$ order. The molecular diameter was set to $\sigma=\widetilde{\tau}=10^{-6}$ in order to maintain the viscosity at relatively low values over orders of magnitude of the reduced density $\eta$, for the comparison with the analytic solution.
The time step was set to $\Delta t=10^{-6}$, the temperature at $T=1$ and the number density and the number density perturbation were set to $\rho_0=0.1$ and $\delta\rho_0=10^{-6}$, respectively. The contact value of pair correlation function $\chi$ is evaluated according to the Standard Enskog theory using $\chi_{\text{\tiny SET}}$ given in Eq.~\eqref{eq:chi}.

The values of the amplitude of the number density $\widetilde{\delta\rho}$ are stored at intervals $T = 100\Delta t$ using the following procedure:
\begin{equation}
 \widetilde{\delta\rho}(t_s)=\frac{2}{N_x}\sum_{i=1}^{N_x} \rho(x_i,t_s)\cos(k x_i)
\end{equation}
where $t_s=s\times T$.

\subsubsection{Simulation results}

Fig.\ref{fig:sound_wave_time_evolution} shows the time evolution of the amplitude $\delta\rho/\delta\rho_0$ obtained using our numerical method, compared with the analytic prediction for the parameters $\alpha_a$, $\alpha_t$ (Eqs.~\eqref{eq:sw_coeff}) and $c_s$(Eq.~\eqref{eq:sw_cs}). Very good agreement can be observed between the numerical and analytic results.

In order to assess the viability of the Simplified Enskog operator, we perform a series of simulations over a couple of orders of magnitude of the reduced density $\eta$. The simulation results are fitted using the analytic solution with the damping coefficients as free parameters and the resulting values are compared to the analytic prediction. The values of the parameters $c_s$, $\alpha_a$ and $\alpha_t$, given by Eqs.~\eqref{eq:sw_cs} and \eqref{eq:sw_coeff}, were obtained using the fitting function given in Eq.~\eqref{eq:sw_solution} and the non-linear least-squares (NLLS) Marquardt-Levenberg fitting algorithm. Fig.~\ref{fig:sound_wave_cs} shows the fitted values of the sound speed $c_s$ compared to the analytic prediction given by Eq.~\eqref{eq:sw_cs}. Excellent agreement is observed throughout the whole span of the reduced density $\eta$.

On the other hand, in the case of the damping coefficients $\alpha_a$ and $\alpha_t$ (Fig.~\ref{fig:sound_wave_alphas}) one can observe very good agreement when the molecular diameter is small enough, i.e. for small values of the reduced density $\eta$. However, the values of these coefficients diverge from the analytic prediction as the reduced density approaches values of $\eta=0.1$ ($E_l=1.10576$). First, the thermal mode $\alpha_t$ is underestimated starting around $\eta=0.1$, while the acoustic mode $\alpha_a$ diverges from the analytic prediction at around $\eta=0.3$ ($E_l=6.3083$). Furthermore, we added in Fig.~\ref{fig:sound_wave_alphas} the analytic values of the damping coefficients in the case of the dilute gas, evaluated according to Eq.~\eqref{eq:sw_coeff} with $P=P_i$ and the shear viscosity given by Eq.~\eqref{eq:viscosity} at the corresponding $\eta$. As expected, the curves converge at small values of $\eta$, as the Enskog collision operator reduces to the Boltzmann one, and at around $\eta=0.01$ the first finite size effects start to appear. This means that the approximation used for the Enskog collision operator works very well for moderately dense gases, but should be applied with care at large values of the reduced density. In the inset of Fig.~~\ref{fig:sound_wave_alphas} we plot the same values but on a linear scale. At the highest value of the reduced density considered ($\eta\approx0.49$) one can observe that the relative error with respect to the analytic prediction in the case of the acoustic mode $\alpha_a$ goes up to around $35\%$.

\subsection{\label{sec:results_shock_wave}Shock wave propagation}

\begin{figure}
  \includegraphics[width=\linewidth]{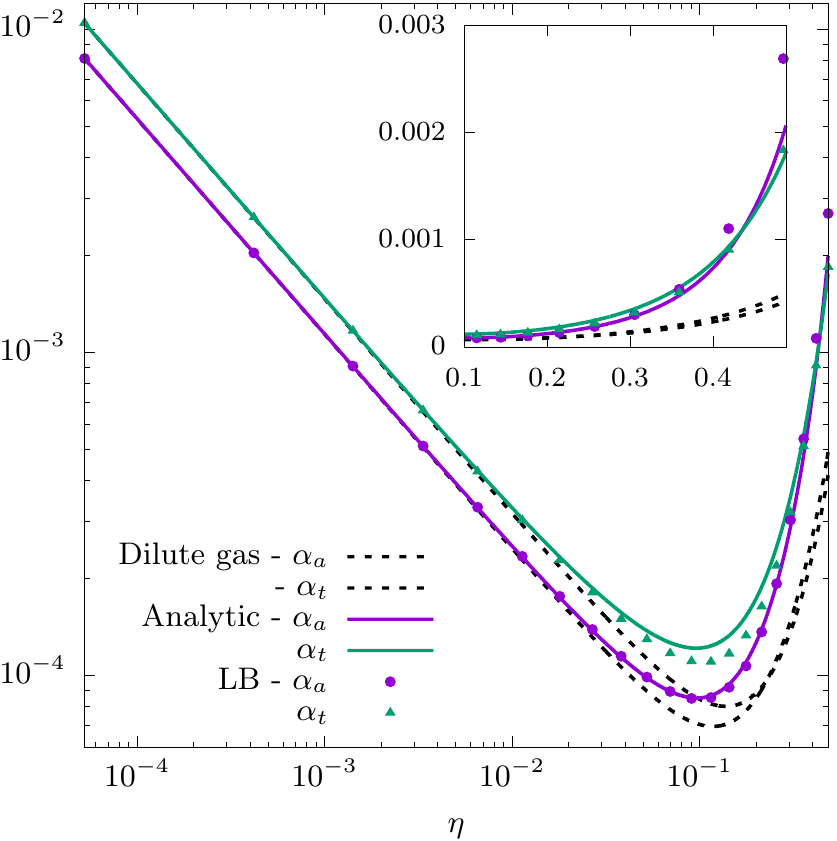}
  \caption{The dependence of both the acoustic $\alpha_a$ and thermal $\alpha_t$ modes with respect to the reduced density $\eta$. The points denote the numerical values obtained using the present model and are fitted using Eq.~\eqref{eq:sw_solution} with $\alpha_a$, $\alpha_t$ and $c_s$ as free parameters. The analytic predictions in Eq.~\eqref{eq:sw_coeff} for the modes $\alpha_a$ and $\alpha_t$ are plotted as solid lines, while their corresponding dilute gas limit is shown as dashed lines. The values of $\alpha_a$ and $\alpha_t$ in the dilute gas limit are evaluated at a viscosity value given by Eq.~\eqref{eq:viscosity} for the corresponding $\eta$. Inset: the same values but on a linear scale. One can observe that the relative error in the case of the acoustic mode $\alpha_a$ goes up to around $35\%$ at the highest value of the reduced density considered($\eta\approx0.49$).}  \label{fig:sound_wave_alphas}
\end{figure}

\begin{figure*}
 	\begin{subfigure}[b]{0.495\textwidth}
		\begin{center}
     \includegraphics[width=\linewidth]{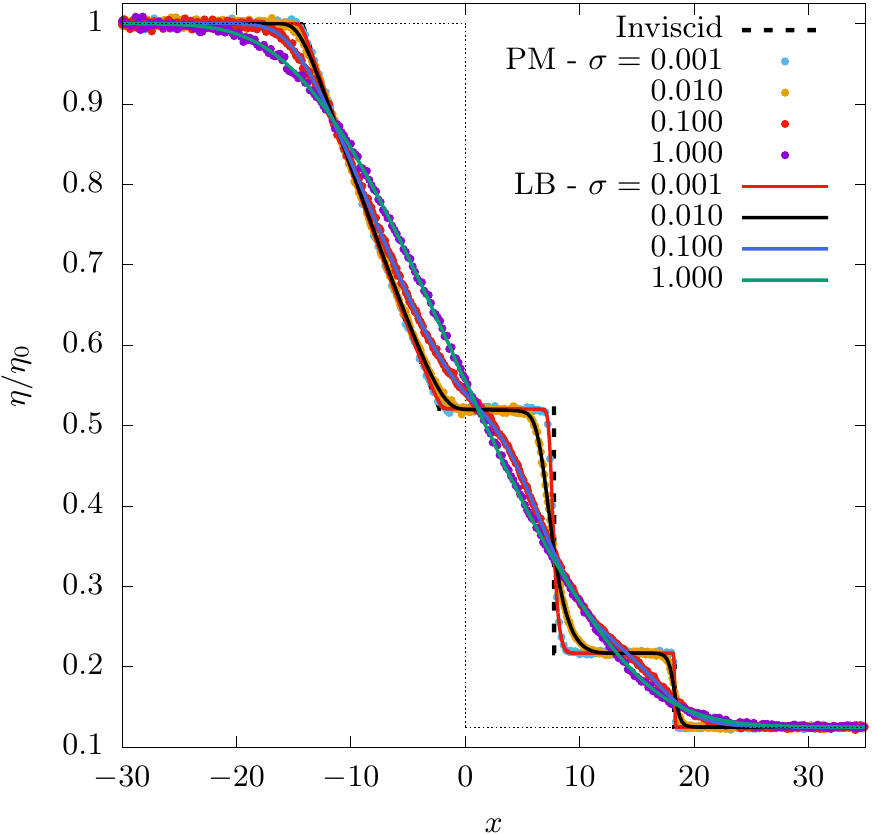}
     \subcaption{Reduced density}
		\end{center}
	\end{subfigure} \hfill
	\begin{subfigure}[b]{0.495\textwidth}
		\begin{center}
			  \includegraphics[width=0.985\linewidth]{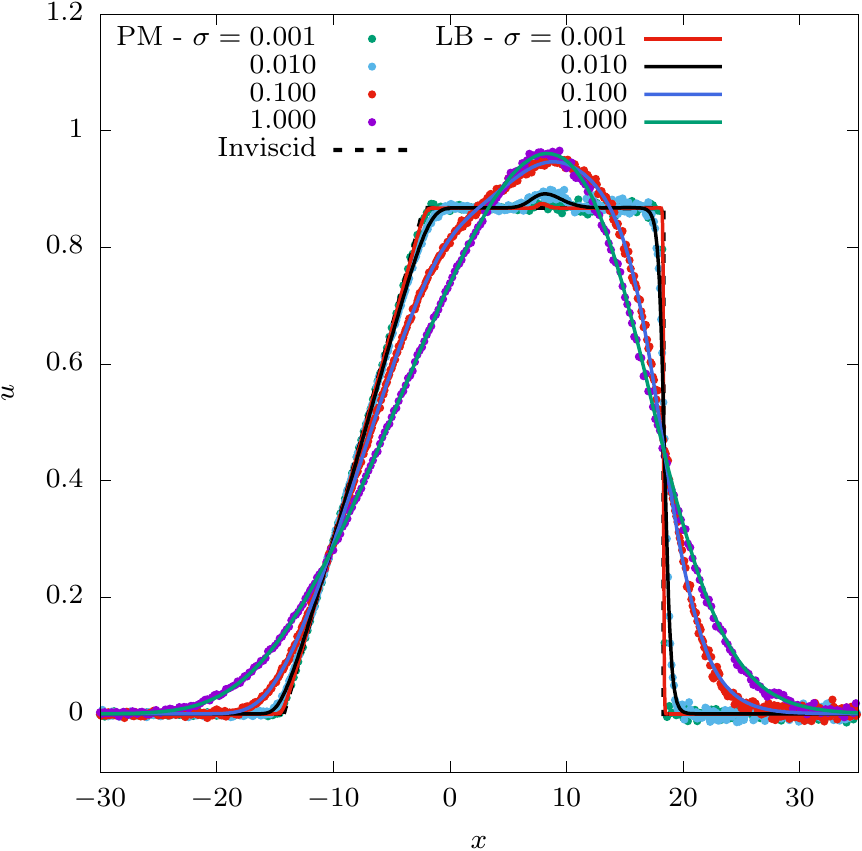}
			\subcaption{Velocity}
		\end{center}
	\end{subfigure}
	 	\begin{subfigure}[b]{0.495\textwidth}
		\begin{center}
      \includegraphics[width=\linewidth]{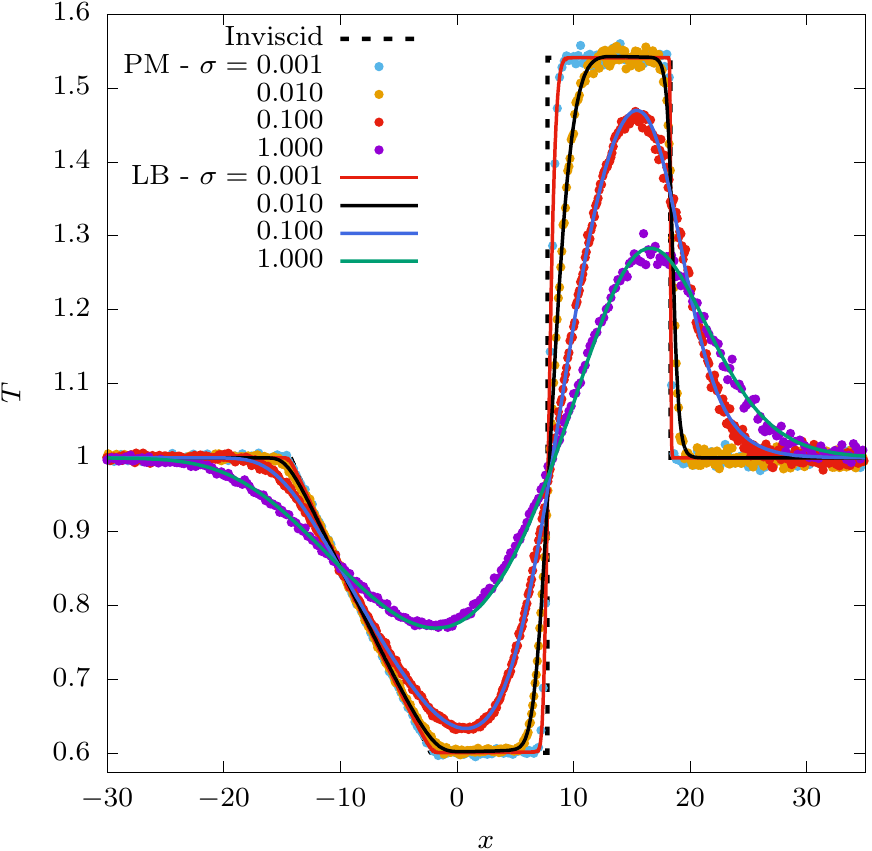}
			\subcaption{Temperature}
		\end{center}
	\end{subfigure}
  \caption{Shock wave propagation. (a) Density, (b) velocity and (c) temperature profiles for constant reduced density$\eta_i=0.05$ ($E_l=0.4825$) but with various values of the molecular diameter (implicitly various values of the relaxation time $\tau$) obtained using the LB model (solid lines) and the particle method PM (points). The dashed line represent the inviscid limit, while in the case of (a) the thin dashed line also shows the initial condition.
  An excellent agreement can be observed for all flow regimes.}
  \label{fig:shock_profiles_na3_001}
\end{figure*}

\begin{figure*}
 	\begin{subfigure}[b]{0.495\textwidth}
		\begin{center}
     \includegraphics[width=0.995\linewidth]{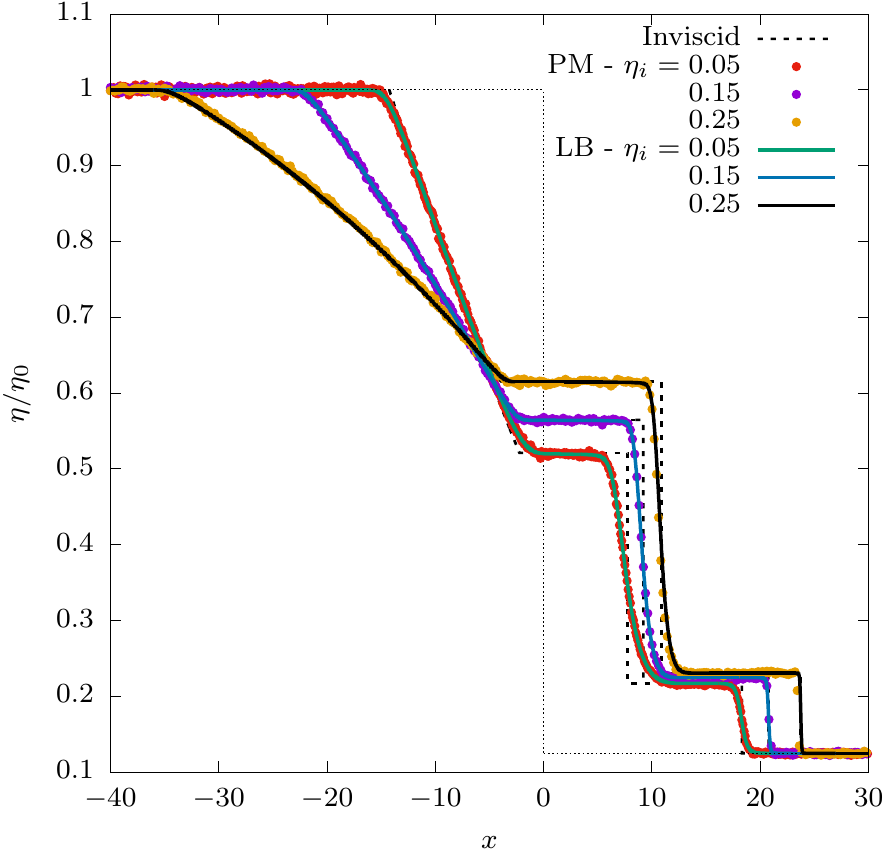}
     \subcaption{Reduced density $\eta/\eta_0$}
		\end{center}
	\end{subfigure} \hfill
	\begin{subfigure}[b]{0.495\textwidth}
		\begin{center}
      \includegraphics[width=\linewidth]{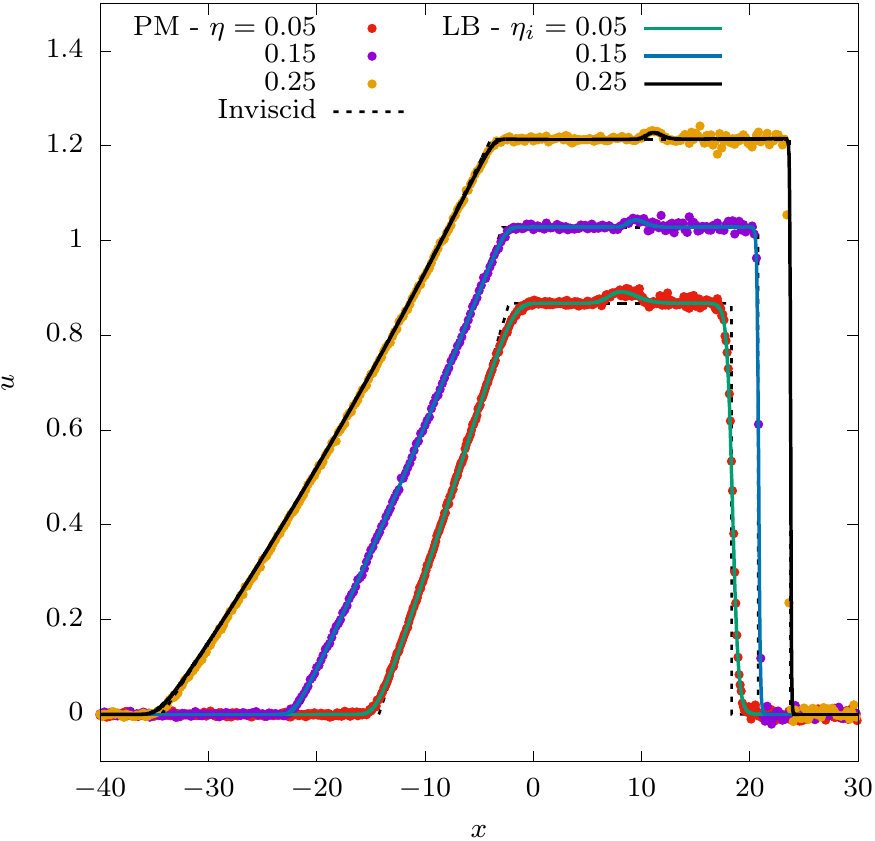}
			\subcaption{Velocity $u$}
		\end{center}
	\end{subfigure}
  \begin{subfigure}[b]{0.495\textwidth}
		\begin{center}
      \includegraphics[width=\linewidth]{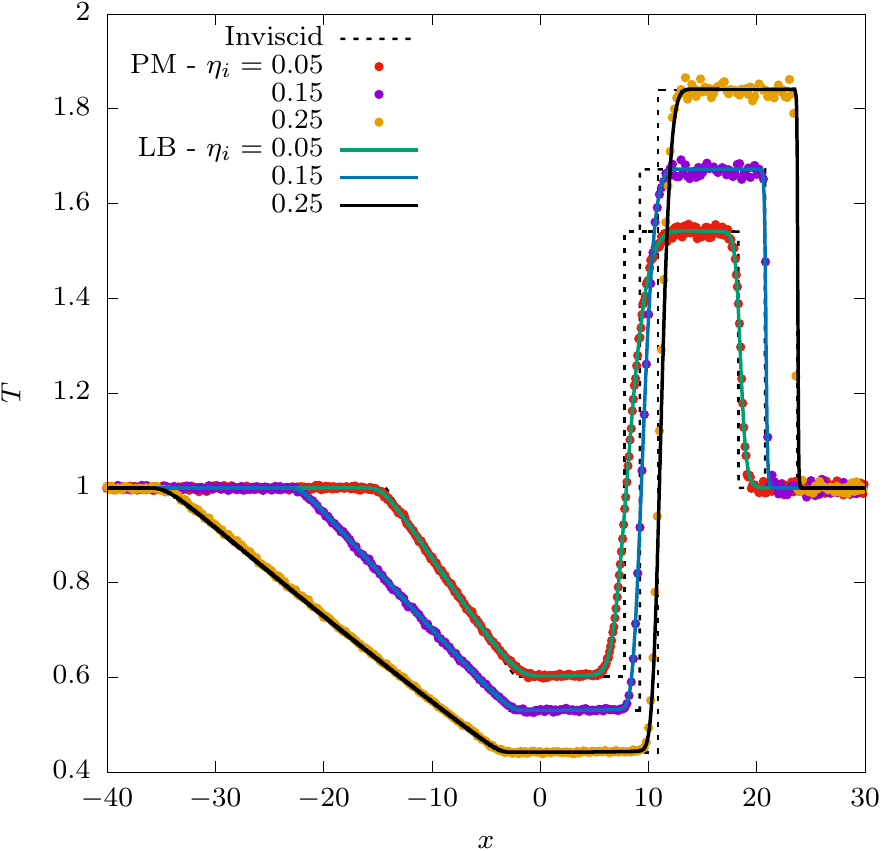}
			\subcaption{Temperature $T$}
		\end{center}
	\end{subfigure}
  \caption{Shock wave propagation. (a) Density, (b) velocity and (c) temperature profiles for constant molecular diameter $\sigma=0.01$ but with various values of the reduced density $\eta\in\{0.0.5,~0.15,~0.25\}$ ($E_l\in\{0.4825,~1.917,~4.3998\}$) obtained using the LB model (solid lines) and the particle method PM (points). The dashed line represent the inviscid limit, while in the case of (a) the thin dashed line also shows the initial condition. Excellent agreement is observed for all values of the reduced density.}
  \label{fig:shock_profiles_sigma001}
\end{figure*}

\subsubsection{Problem statement: 1D Sob shock tube}

The 1D Sod shock tube problem was proposed by G. A. Sod in 1978\cite{S78}. Consider a membrane located at $x = x_0$ that separates two semi-infinite domains. The fluid properties are homogeneous in each domain, while the velocity is zero everywhere. At the initial time, the fluid properties are:
\begingroup
\renewcommand*{\arraystretch}{1.25}
\begin{equation}\label{eq:shock_init}
 \begin{pmatrix}
 \eta_L \\ T_L \\u_L
 \end{pmatrix}=
\begin{pmatrix}
 \eta_i \\ 1.0 \\ 0.0
\end{pmatrix},\quad
 \begin{pmatrix}
 \eta_R \\ T_R \\ u_R
 \end{pmatrix}=
\begin{pmatrix}
 \eta_i/8 \\ 1.0 \\ 0.0
\end{pmatrix}
\end{equation}
\endgroup
where $\eta_i$ is the initial value of the reduced density in the left domain.

\subsubsection{\label{sec:inviscid_limit}Inviscid limit}

We describe here the standard approach to the solution in the inviscid regime found in many textbooks\cite{F95,KCD15,W16} and adapt it to the case of the dense gas.

Starting from the Euler equations:
\begin{subequations}\label{eq:euler}
\begin{align}
 \frac{D\rho}{Dt}+\rho\nabla \bm{u}&=0\label{eq:euler_mass}\\
 \rho \frac{D\bm{u}}{Dt}+\nabla P&=0\label{eq:euler_momentum}\\
 \rho\frac{De}{Dt}+P\nabla\bm{u}&=0\label{eq:euler_energy}
 \end{align}
\end{subequations}
one can introduce the similarity variable:
\begin{equation}
 \xi=\frac{x-x_0}{t}.
\end{equation}
In this case the Eqs.~\eqref{eq:euler} reduce to:
\begin{subequations}\label{eq:euler_xi}
\begin{align}
\partial_\xi u-\frac{\xi-u}{\rho}\partial_\xi\rho&=0\\
\partial_\xi P-(\xi-u)^2\partial_\xi\rho&=0
 \end{align}
\end{subequations}

By replacing the above equations in Eq.~\eqref{eq:euler_energy} and assuming that $\partial_\xi\rho\neq0$, the equations are satisfied either when $u=\xi$, corresponding to the contact discontinuity, or when:
\begin{equation}\label{eq:sod_sol}
 u=\xi\pm c_s
\end{equation}
The ($+$) solution refers to the rarefaction head, travelling to the left, while the ($-$) solution is the rarefaction tail.
Since at the head of the rarefaction wave $u=u_L=0$, the velocity of the head is constant and is given by:
\begin{equation}
 \xi_r=-c_s
\end{equation}
while the tail of the rarefaction wave travels with the constant value on the plateau $u=u_c$:
\begin{equation}
 \xi_c=u_c-c_s
\end{equation}

Replacing Eq.~\eqref{eq:sod_sol} in Eqs.~\eqref{eq:euler_xi}, one obtains the system of equations for the rarefaction wave:
\begin{subequations}
 \begin{align}
 1 + \frac{1}{2c_s}\left(\partial_\rho c_s^2 \partial_\xi\rho + \partial_P c_s^2\partial_\xi P\right)&=-c_s\partial_\xi\ln\rho\\
 \partial_\xi P&=c_s^2\partial_\xi\rho
 \end{align}
\end{subequations}
where the sound speed in Eq.~\eqref{eq:sw_cs} is written in terms of $\rho$ and $P$ as:
\begin{equation}
 c_s^2(\rho,P)=\frac{P}{\rho}+\frac{2P(1+b\rho\chi)}{3\rho}+\frac{bP(\chi+\rho\partial_\rho\chi)}{1+b\rho\chi}
\end{equation}

This system of equations can be solved numerically in conjunction with the Rankine-Hugoniot relations for the discontinuity (i.e. shock front) travelling with velocity $\xi_s$, given by:
\begin{subequations}\label{eq:rankine_hugoniot}
\begin{align}
\rho_2(u_c-\xi_s)&=-\xi_s\rho_R \\
\rho_2 u_c (u_c-\xi_s)+P_c&=P_R\\
(e_c+\frac{1}{2}\rho_2 u_c^2)(u_c-\xi_s)+u_c P_c &= e_R\xi_s
 \end{align}
\end{subequations}
where the following notations have been introduced:
\begin{multline}
\rho_1=\rho(\xi_c),\,\rho_2=\rho(\xi_s),\,e_c=e(\rho_c,T_c),\,e_R=e(\rho_R,T_R)\\ P_c=P(\rho_1,T_1)=P(\rho_2,T_2),\, P_R=P(\rho_R,T_R)
\end{multline}
where the subscript $1$ and $2$ refer to the left and right side of the contact discontinuity.

The solution is obtained using the high-precision numerical solver included in the software package Mathematica\textregistered\cite{Mathematica}.

\subsubsection{Computational setup}

The simulations are performed on a system of length $L=80$ and temperature $T=1$. The contact value of pair correlation function $\chi$ is evaluated according to the revised Enskog theory using $\chi_{\text{\tiny RET-FM}}$ given in Eq.~\eqref{eq:chi_ret}.

\paragraph{Lattice Boltzmann}

The number of nodes varies depending on the molecular diameter $\sigma$, from $N_x=8\times10^3$ at $\sigma=10^{-3}$ to $N_x=160$ at $\sigma=1$. The large number of nodes at small $\sigma$ is made equal to the number of computational cells in the particle method and it offers sufficient resolution to reveal the features of the shock wave. The quadrature order is set to $Q_x=8$ for $\sigma<0.1$, $Q_x=20$ for $\sigma=0.1$, while for $\sigma=1$ a quadrature of $Q_x=200$ was necessary since the flow is close to the ballistic regime. The time step was set at $\Delta t=10^{-3}$.

\paragraph{Particle method}

The results for the particle method are obtained by averaging over 10 runs comprised of $N_p=1.6\times10^{7}$ particle per run, in a system of $8\times10^{3}$ computational cells. Also in this case, the time step was set to $\Delta t=10^{-3}$.

\subsubsection{Numerical results}

In this subsection, we compare the results obtained using the Lattice Boltzmann implementation versus the results obtained using the particle method presented in Sec.~\ref{sec:ensk_dsmc}.

\paragraph{\label{sec:shock_t9}Shock profiles at various relaxation times}

\begin{figure*}
 	\begin{subfigure}[b]{0.33\textwidth}
		\begin{center}
     \includegraphics[width=\linewidth]{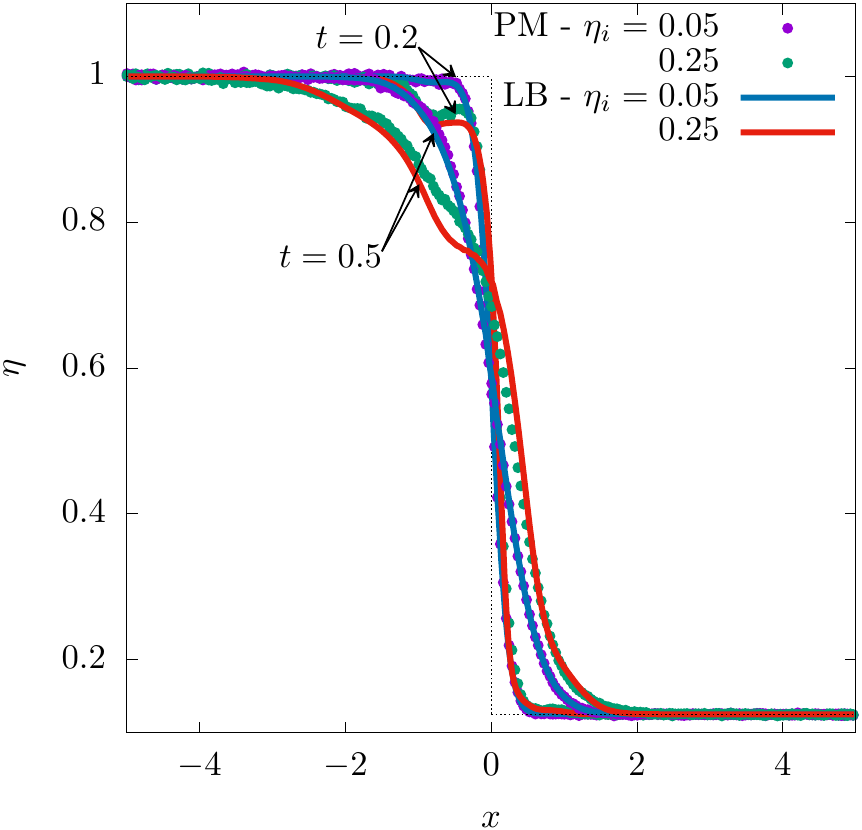}
     \subcaption{Reduced density $\eta/\eta_0$}
		\end{center}
	\end{subfigure} \hfill
	\begin{subfigure}[b]{0.33\textwidth}
		\begin{center}
      \includegraphics[width=\linewidth]{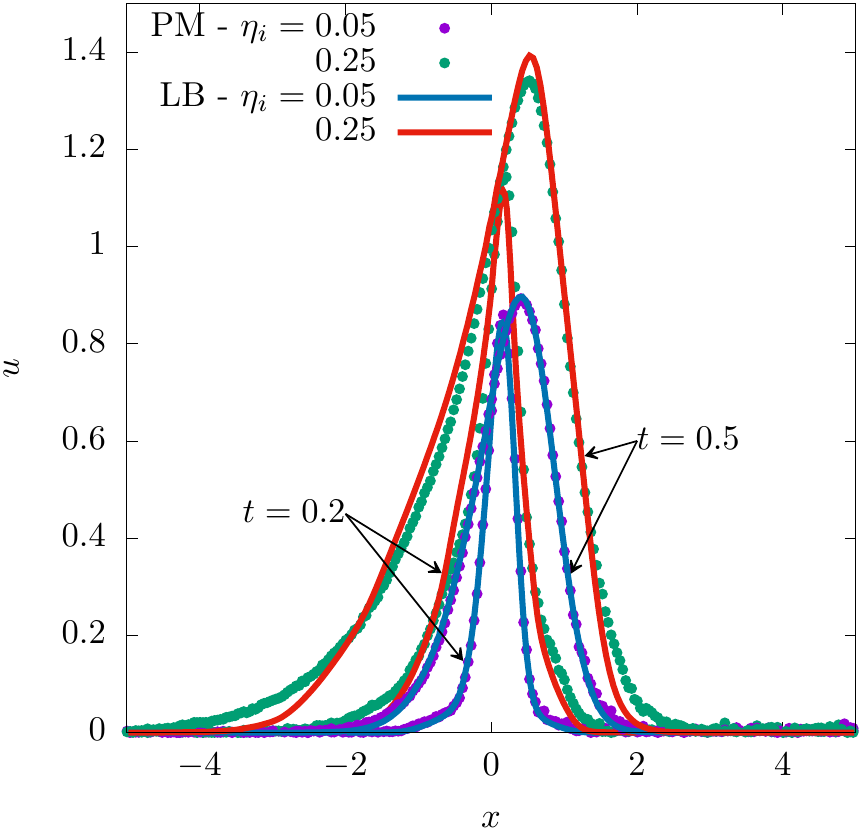}
			\subcaption{Velocity $u$}
		\end{center}
	\end{subfigure}\hfill
  \begin{subfigure}[b]{0.33\textwidth}
		\begin{center}
      \includegraphics[width=\linewidth]{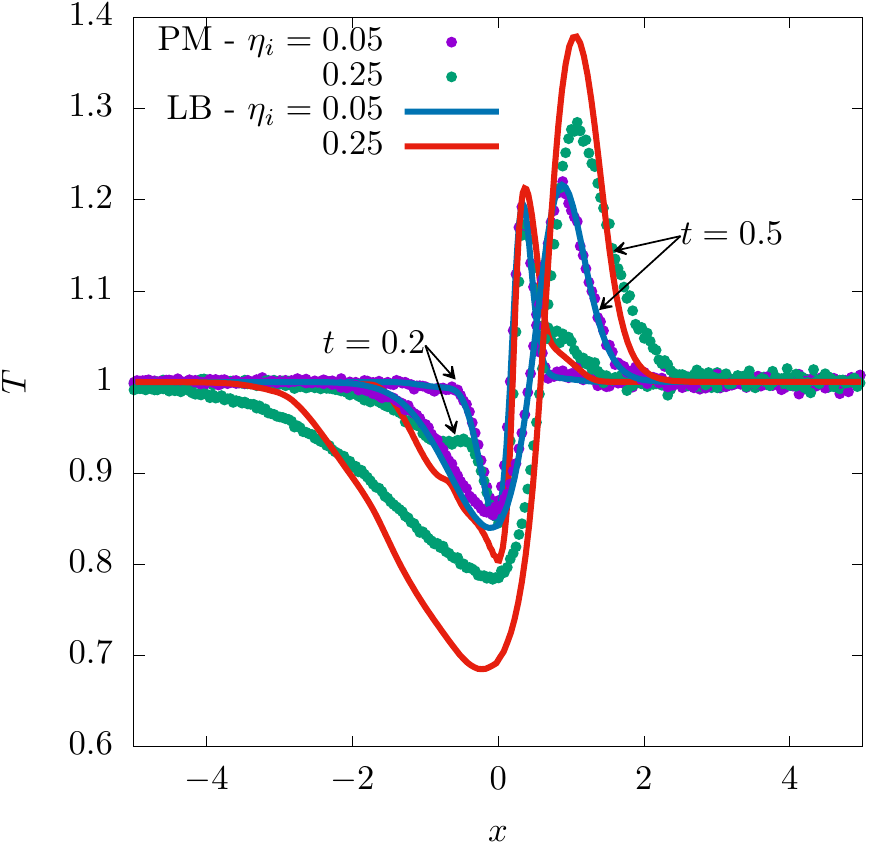}
			\subcaption{Temperature $T$}
		\end{center}
	\end{subfigure}
  \caption{Shock wave propagation: structure at the initial time. (a) Density, (b) velocity and (c) temperature profiles for molecular diameter $\sigma=1$ at reduced density $\eta_i=\{0.05,0.25\}$ ($E_l\in\{0.4825,4.3998\}$, respectively) obtained using the LB model (solid lines) with quadrature $Q_x=200$ and the PM method (points), at two time instances $t\in\{0.2,0.5\}$. In the case of (a) the thin dashed line shows the initial condition.}
  \label{fig:shock_profiles_ret}
\end{figure*}

At first, we will consider the initial conditions listed in Eq.~\eqref{eq:shock_init}. Fig.~\ref{fig:shock_profiles_na3_001} present the results for 4 values of the molecular diameter $\sigma=\{10^{-3},10^{-2},10^{-1},10^{0}\}$, while keeping the reduced density $\eta_i=0.05$ constant, resulting in 4 different relaxation times $\tau$, in a system of length $L=80$. This relatively large size of the system is required due to the high computational costs associated with the particle method at small values of the molecular diameter $\sigma$.
The profiles of reduced density $\eta$, velocity $u$ and temperature $T$ are presented alongside the inviscid limit. Very good agreement can be observed for all flow regimes, from hydrodynamic to the near ballistic regime. The LB results are plotted using solid lines, the particle method results are represented by solid circles and the dashed line represents the inviscid limit obtained by numerically solving the equations in Sec.~\ref{sec:inviscid_limit}. Please refer to Sec.~\ref{sec:inviscid} for further results close to the inviscid regime obtained at $\sigma=10^{-6}$ and $\eta_i=0.05$ ($E_l=0.4825$), and Sec.~\ref{sec:ballistic} for details about the choice of quadrature at the near ballistic regime ($\sigma=1$).

Next, we fixed the molecular diameter at $\sigma=0.01$ and varied the reduced density $\eta$. Due to the high computational demand of the PM, scaling with the particle number density, we have chosen the above value of the molecular diameter since it is small enough to be compared to the inviscid limit. The set of reduced densities on the left-hand side is $\eta_i=\{ 0.05,0.15,0.25 \}$ ($E_l\in\{0.4825,~1.917,~4.3998\}$). Very good agreement between the LB and PM results is observed for all values of the initial reduced density $\eta$, as well as for each considered macroscopic quantity, namely the reduced density $\eta$, the velocity $u$ and the temperature $T$.

In terms of computational time, it is expected that the LB method is much faster than the PM. This is expressed quantitatively in Table~\ref{tab:comp_time}, where the running times for each method, namely $t_{\text{\tiny LB}}$ and $t_{\text{\tiny PM}}$, are evaluated using a single core of an Intel\textregistered{} Xeon\textregistered{} Gold 6330 CPU. The time ratio $t_{\text{\tiny PM}}/t_{\text{\tiny LB}}$ varies from $10^4$ at $\sigma=0.001$ and $120$ at $\sigma=1$. As it can be seen in the table the running times for the LB increase with $\sigma$ due to the larger velocity set needed, while for the PM the number of collisions scales with the inverse of the molecular diameter ($N_c\approx1/\sigma$). As expected, the ratio $t_{\text{\tiny PM}}/t_{\text{\tiny LB}}$ increases for smaller relaxation time $\tau$ (at constant reduced density $\eta$ the relaxation time is proportional to the molecular diameter $\sigma$). The listed times for the PM method are for only one run, a series of 10 runs have been executed to obtain the results presented in Fig.~\ref{fig:shock_profiles_na3_001}.

\begin{table}
	\begin{center}
		\begin{tabular*}{\columnwidth}{@{\extracolsep{\stretch{1}}}*{7}{l|ccc|c|c}@{}}
			\toprule
			  Method$\rightarrow$ & LB & & & PM & \\ \hline
       $\sigma$    & $Q_x$ & $N_x$  & $t_{\text{\tiny LB}}$  & $t_{\text{\tiny PM}}$    & $t_{\text{\tiny PM}}/t_{\text{\tiny LB}}$ \\ \hline
			 0.001 &  8   & 1600 & 62s &  186h  & $\approx 1.1\times10^4$ \\ \hline
			 0.01  &  8   & 800 & 32s &  23h   & $\approx 2.5\times10^3$ \\ \hline
       0.1  &  20  & 640 & 71s  &  7.25h & $\approx 370$\\ \hline
			 1   &  200  & 160 & 176s  &  5.8h & $\approx 120$\\
			 \hline \hline
		\end{tabular*}
	\end{center}
	\caption{Computational time comparison for the simulations presented in Fig.~\ref{fig:shock_profiles_na3_001}. As expected, the ratio $t_{\text{\tiny PM}}/t_{\text{\tiny LB}}$ increases for smaller relaxation time $\tau$, since at constant reduced density $\eta$ the relaxation time is proportional to the molecular diameter $\sigma$ (Eq.~\eqref{eq:tau_sigma}).}
	\label{tab:comp_time}
\end{table}

\paragraph{Shock structure at the initial times}

At first glance, the shock profiles presented in the above section look qualitatively similar to the shock profiles for dilute gases, comprised of a rarefaction wave, the two plateaus separated by the contact discontinuity and the shock front. However, in their initial stage (i.e. close to the ballistic regime, due to the self-similarity of the shock), the dense gas shock wave deviates from the shape for dilute gases, at length scales comparable to the molecular diameter $\sigma$. More precisely, the discrepancies become negligible at the scales used in Sec.~\ref{sec:shock_t9}, as the system length is much larger than the molecular diameter.
Here we present the results for the shock profiles at $t=\{0.2,0.5\}$ for $\sigma=1$ and for two values of the reduced density $\eta_i\in\{0.05,0.25\}$ ($E_l\in\{0.4825,4.3998\}$, respectively). To obtain these results, we employed the quadrature order $Q_x=200$ and $N_x=400$ nodes at $\eta_i=0.05$ and $N_x=200$ nodes at $\eta_i=0.25$, in a system of length $L=10$.

At first, the density develops a quasi plateau that is dissipated relatively fast, in contrast with the ballistic results, due to the nonlocal interactions. At low reduced density $\eta$, the LB model is able to reproduce quantitatively and qualitatively the PM results for all macroscopic quantities, while at large $\eta$ the features of the shock are recovered only qualitatively, discrepancies been observed for all macroscopic quantities, especially the temperature. This further denotes that the approximation used for the Enskog collision integral gives good accuracy up to a moderate value of the reduced density $\eta$.

\paragraph{\label{sec:inviscid}Inviscid regime}

\begin{figure}
 	\begin{subfigure}[b]{0.475\textwidth}
		\begin{center}
     \includegraphics[width=0.975\linewidth]{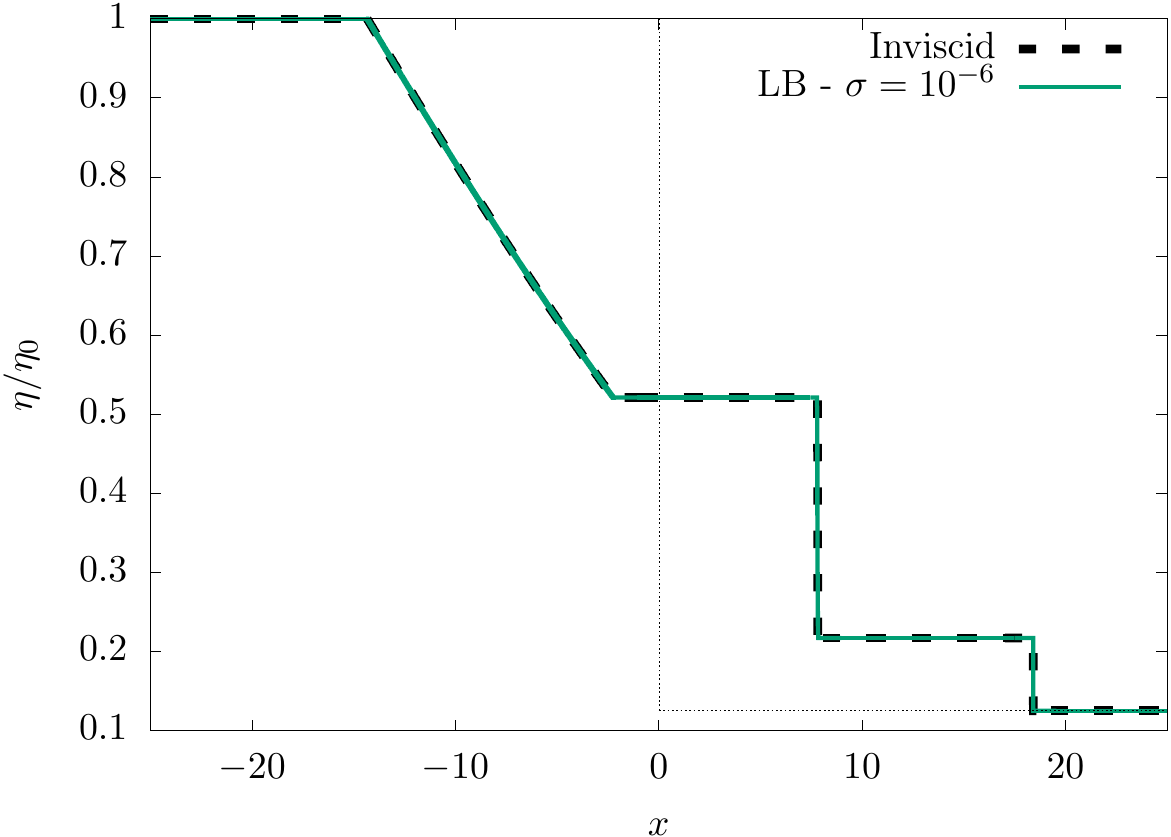}
     \subcaption{Reduced density $\eta/\eta_0$}
		\end{center}
	\end{subfigure} \hfill\\
	\begin{subfigure}[b]{0.475\textwidth}
		\begin{center}
      \includegraphics[width=0.975\linewidth]{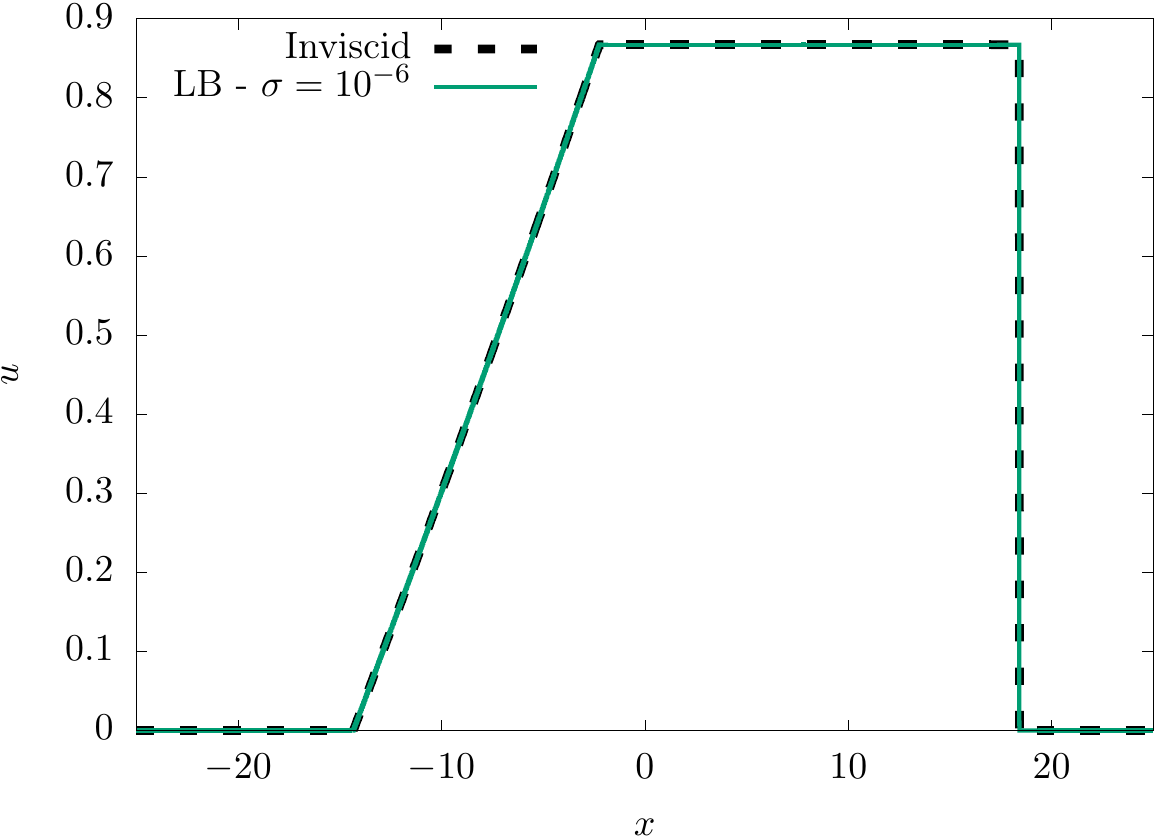}
			\subcaption{Velocity $u$}
		\end{center}
	\end{subfigure}\hfill\\
  \begin{subfigure}[b]{0.475\textwidth}
		\begin{center}
      \includegraphics[width=0.975\linewidth]{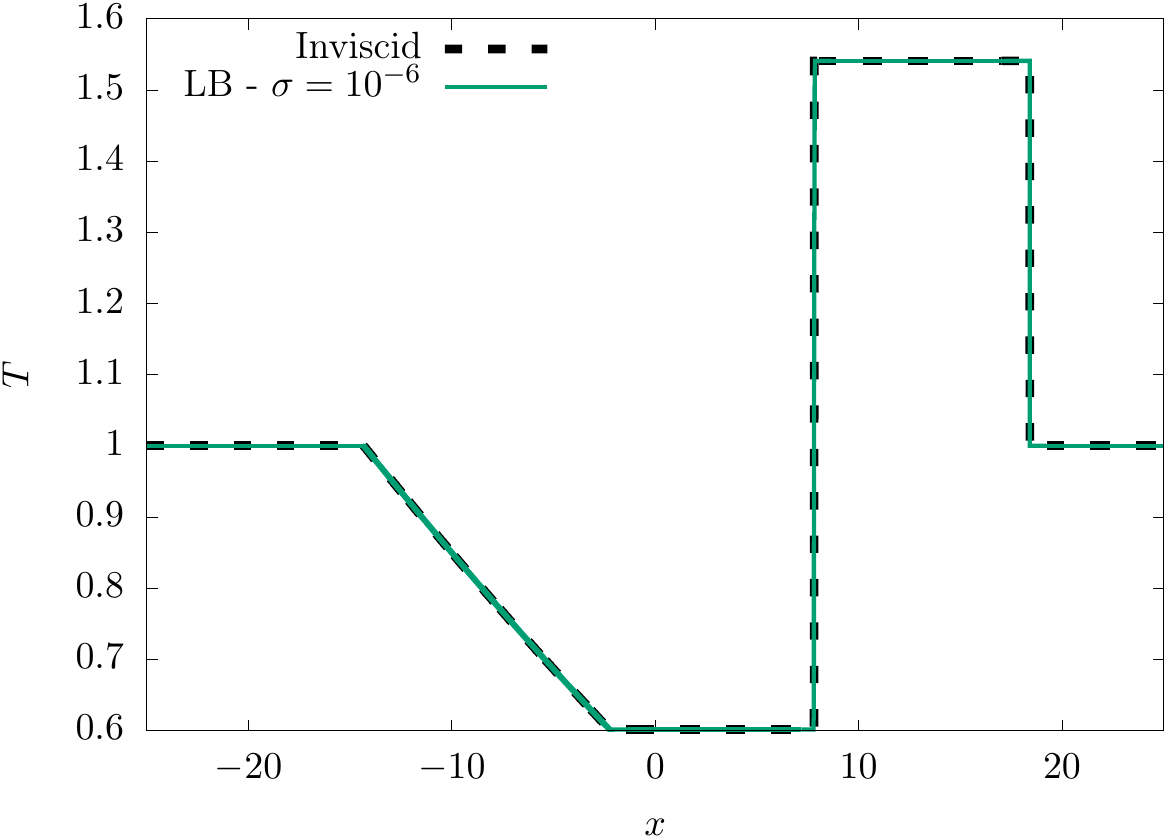}
			\subcaption{Temperature $T$}
		\end{center}
	\end{subfigure}
  \caption{Shock wave propagation. (a) Density, (b) velocity and (c) temperature profiles for molecular diameter $\sigma=10^{-6}$ at reduced density $\eta_i=0.05$ ($E_l=0.4825$) obtained using the LB model (solid line) and compared with the inviscid solution (dashed line). In the case of (a) the thin dashed line shows the initial condition. Perfect overlap can be observed for all macroscopic quantities. In the case of (a) the thin dashed line also shows the initial condition.
  }
  \label{fig:shock_profiles_sigma10-6}
\end{figure}

The results for the near inviscid regime were obtained using the proposed LB model in a system of length $L=80$. The relaxation scaling factor was set to $\widetilde{\tau}=10^{-6}$ and the reduced density was set to $\eta_i=0.05$ ($E_l=0.4825$), being equivalent to a molecular diameter of $\sigma=10^{-6}$ and $\chi_{\text{\tiny SET}}$ was used. For better resolution at sharp interfaces, a number of $N_x=8\times10^{4}$ nodes have been used ($\Delta x=10^{-3}$), and a time-step of $\Delta t=10^{-6}$. The results are plotted in Fig.~\ref{fig:shock_profiles_sigma10-6} and one can observe a perfect overlap between the analytic solution and the LB results.

\paragraph{\label{sec:ballistic}Near ballistic regime}

\begin{figure*}
 	\begin{subfigure}[b]{0.33\textwidth}
		\begin{center}
     \includegraphics[width=\linewidth]{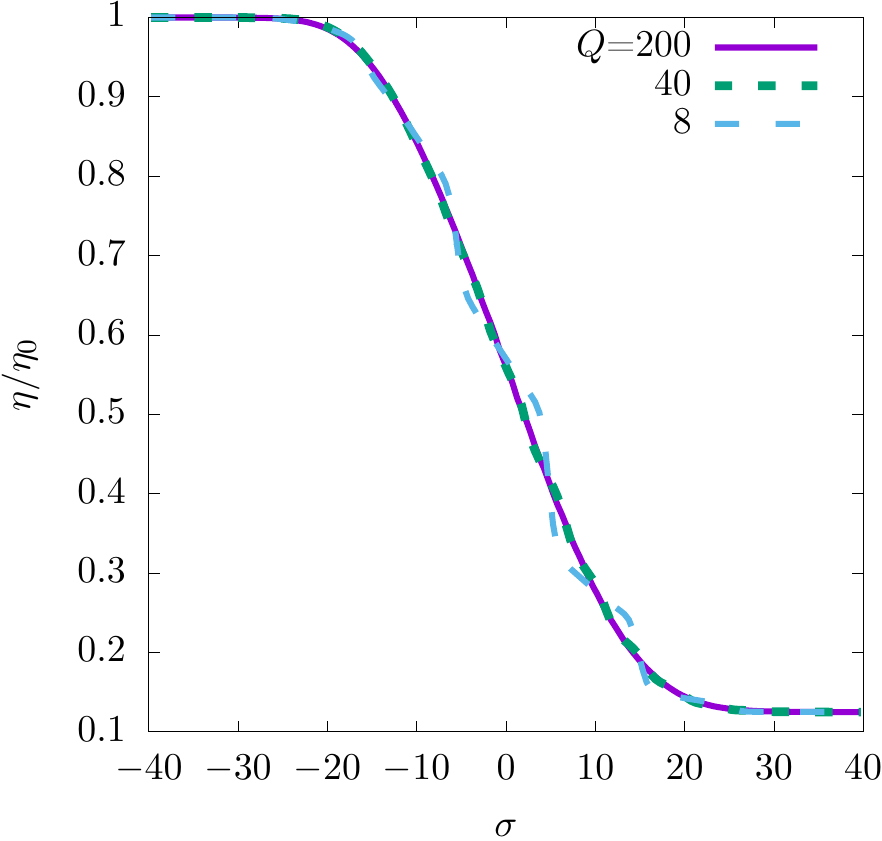}
     \subcaption{Reduced density $\eta/\eta_0$}
		\end{center}
	\end{subfigure} \hfill
	\begin{subfigure}[b]{0.33\textwidth}
		\begin{center}
      \includegraphics[width=\linewidth]{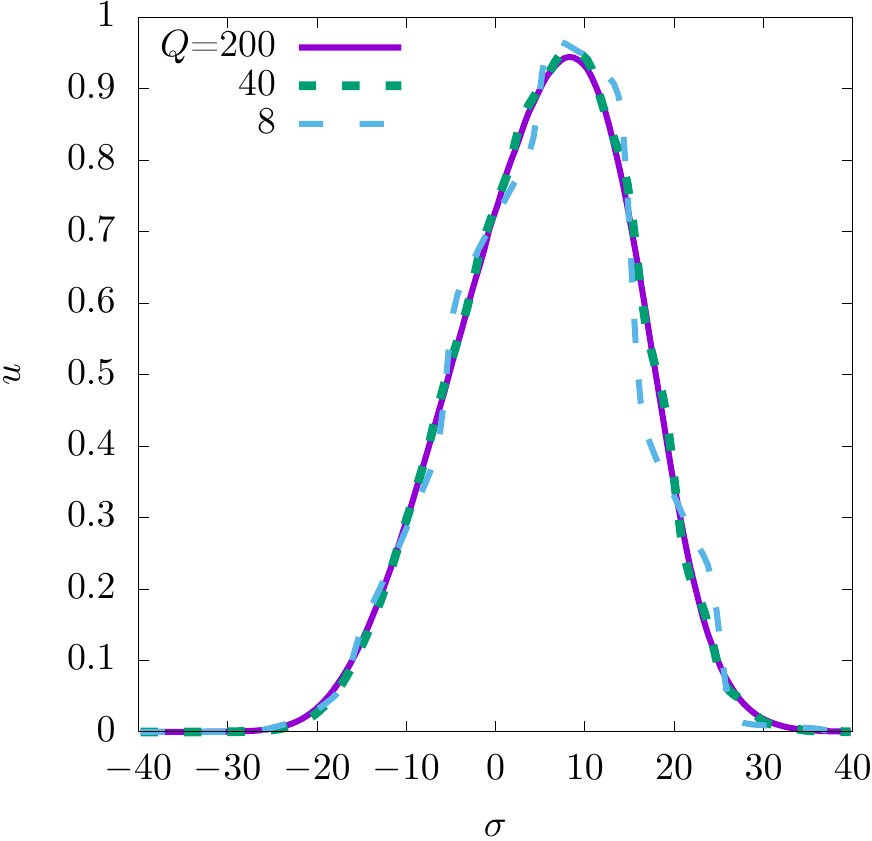}
			\subcaption{Velocity $u$}
		\end{center}
	\end{subfigure}\hfill
  \begin{subfigure}[b]{0.33\textwidth}
		\begin{center}
      \includegraphics[width=\linewidth]{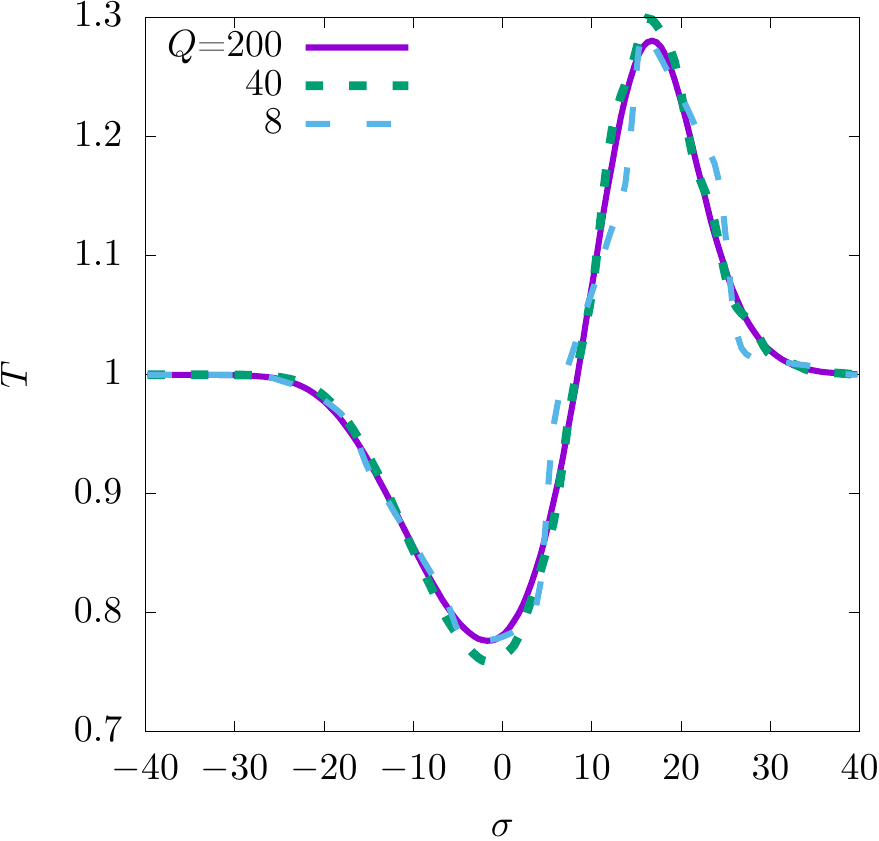}
			\subcaption{Temperature $T$}
		\end{center}
	\end{subfigure}
  \caption{Shock wave propagation. (a) Density, (b) velocity and (c) temperature profiles for molecular diameter $\sigma=1$ at reduced density$\eta_i=0.05$ ($E_l=0.4825$) obtained using the LB model (solid lines) with quadrature $Q_x=\{8,40,200\}$. In the near ballistic regime, one needs to employ a large velocity set in order to smooth out the profiles.}
  \label{fig:shock_profiles_ballistic}
\end{figure*}

As the relaxation time is growing, a larger momentum space is needed in order to capture the collisionless behaviour. A small number of velocities would render a staircase solution since collisions play a very small role in particle evolution. As such, a large number of momentum points need to be employed in order to obtain a smooth profile of the macroscopic quantities. In all simulations, the molecular diameter is set to $\sigma=1$ at reduced density$\eta_i=0.05$ ($E_l=0.4825$) and the number of nodes is $N_x=160$ and the time step is $\Delta t =10^{-3}$. In Fig.~\ref{fig:shock_profiles_ballistic} we present the profiles of reduced density, velocity and temperature at three values of the quadrature order $Q_x\in{8,40,200}$, chosen to have a factor of 5 between them in order to track the improvement of the profiles. One can observe that at $Q=200$ the profiles are smooth enough and they agree very well with the PM results, as presented in Fig.~\ref{fig:shock_profiles_na3_001}.

\section{\label{sec:conclusions} Conclusions}

In this work, the propagation of longitudinal, as well as of the shock waves in dense gases are simulated in order to validate the proposed finite-difference Lattice Boltzmann model employing the simplified Enskog collision integral. In this model, the Enskog collision integral is approximated using a Taylor expansion and retaining the first-order gradients. The simulation results for the longitudinal waves were compared to the analytic solution for various values of the reduced density $\eta$. The simulation results for shock waves were compared to the results obtained using a particle method for the solution of the Enskog equation.

The sound wave propagation was used to check the applicability domain of the simplified Enskog collision operator with respect to the reduced density $\eta$. The sound speed values are accurately recovered in the LB simulations, while the damping coefficients show deviations from the analytic prediction as the reduced density $\eta$ is increased. We observed that the discrepancies appear around $\eta=0.1$ ($E_l=1.10576$) and become significant at $\eta=0.3$ ($E_l=6.3083$). Beyond these values, one can still obtain results with reasonable accuracy using the simplified Enskog collision operator. Higher-order terms might be needed in order to extend the applicability of the present model.

Shock wave propagation is employed to test the capabilities of the numerical schemes when sharp variations in macroscopic quantities are present. The results are compared with a particle method that solves the Enskog collision integral using a Monte-Carlo method.
The simulations were conducted for various values of the relaxation time $\tau$, as well as various values of the reduced density.
For large systems and small values of the relaxation time $\tau$ (i.e. small molecular diameters with respect to the system extension) the LB results overlap very well with the PM results for a good range of reduced densities values $\eta\in\{0.05,0.15,0.25\}$. When looking at the initial stages of the shock wave propagation, at scales comparable to the molecular diameter, one can observe some features that are not present in the dilute gas regime. These features are well captured by the LB model at small values of the reduced density ($\eta=0.05$), while at large values ($\eta=0.25$) the discrepancies are significant. We also presented that the scheme can recover the inviscid regime with a perfect overlap over the analytic solution. The overlap between the two methods is remarkably good, given the huge computational time difference between the two methods ($2$ to $4$ orders of magnitude).

We conclude that this model is able to deal with moderately dense gases. Moreover, we determined the applicability range of the simplified Enskog collision operator and challenged the proposed model in tackling flows with sharp gradients in the macroscopic quantities. In the future, we plan to consider also gas-surface interactions and as well as to introduce attractive forces between molecules, in order to tackle bounded flows and multiphase flows, respectively.

\begin{acknowledgments}
The authors thank V.E. Ambrus and V. Sofonea for useful discussions regarding the present manuscript. This work was supported through a grant from the Ministry of Research, Innovation and Digitization, CNCS - UEFISCDI, project number PN-III-P1-1.1-PD-2021-0216, within PNCDI III.

\end{acknowledgments}

\section*{Data Availability Statement}

The data that support the findings of this study are available from the corresponding author upon reasonable request.

\section*{Author declarations}
\subsection{Conflict of Interest}
The authors have no conflicts to disclose.

\appendix

\section{\label{app:numerical_scheme}Numerical schemes for the LB implementation}


\subsection{Third-order TVD Runge-Kutta method}

In order to implement the time-stepping algorithm, it is convenient to cast the Boltzmann equation \eqref{eq:evolution} in the following form:
\begin{equation}\label{eq:cast}
 \partial_t f_{k} = L[f_{k}], \qquad
 L[f_{k}] = - \frac{{p}_{k}}{\,m\,}\cdot \nabla f_{k}
-\frac{1}{\tau}[f_{\bm{\kappa}} - f^{S}_{\bm{\kappa}}]+J^1_{k}.
 \end{equation}

The third-order total variation diminishing (TVD) Runge-Kutta integrator gives the following three-step algorithm for computing the values of $f_{k}$ at time $t+ \delta t$\cite{SO88,GS98,RZ13}:
\begin{align}
 f_{k}^{(1)}(t) =& f_{k}(t) + \delta t \, L[f_{k}(t)], \nonumber\\
 f_{k}^{(2)}(t) =& \frac{3}{4} f_{k}(t) + \frac{1}{4} f_{k}^{(1)}(t) +
 \frac{1}{4} \delta t\, L[f_{k}^{(1)}(t)],\nonumber\\
 f_{k}(t+\delta t)
 =& \frac{1}{3} f_{k}(t) + \frac{2}{3} f_{k}^{(2)}(t) +
 \frac{2}{3} \delta t\, L[f_{k}^{(2)}(t)]. \label{eq:rk3}
\end{align}

The Butcher tableau \cite{B08} corresponding to this scheme is given in Table~\ref{tab:rk3}.

\begin{table}[b]
\begin{center}
\caption{Butcher tableau associated with the third-order Runge-Kutta
time-stepping procedure described in Eq.~\eqref{eq:rk3}.\label{tab:rk3}}
\begin{tabular}{r|rrr}
0 & & & \\
1 & 1 & & \\
1/2 & 1/4 & 1/4 & \\\hline
& 1/6 & 1/6 & 2/3
\end{tabular}
\end{center}
\end{table}

\subsection{WENO-5 advection scheme}

The advection term which appears in Eq.~\eqref{eq:cast} above, namely $p_{k}\cdot \nabla f_{k}/m$ is computed using the Weighted Essentially Non-Oscillatory scheme of order $5$ (WENO-5) along each coordinate\cite{GXZL11,JS96}. We will describe in the following the one-dimensional case. Assuming that the flow domain is discretized using $1 \le i \le N$ nodes on the $x$ axis, the advection term becomes:
\begin{equation}
 \left(\frac{p_{k}}{m} \cdot \partial_x f_{k}\right)_{k; i} =
 \frac{\mathcal{F}_{k; i + 1/2} - \mathcal{F}_{k; i - 1/2}}
 {\delta s}
\end{equation}
where $\mathcal{F}_{k;i+1/2}$ represents the flux of $f$ advected with velocity $p_{k} / m$ through the interface between the cells centered on $\bm{x}_{i}$ and $\bm{x}_{i+1}$. The construction of these fluxes is summarized below, under the assumption of a positive advection velocity $p_{k} / m > 0$. In this case, the flux $\mathcal{F}_{k; i+1/2}$ can be computed using the following expression\cite{GXZL11}:
\begin{equation}\label{eq:weno5_flux_x}
\mathcal{F}_{i+1/2} = \overline{\omega}_1\mathcal{F}^1_{i+1/2} +
\overline{\omega}_2\mathcal{F}^2_{i+1/2} + \overline{\omega}_3\mathcal{F}^3_{i+1/2},
\end{equation}
where for brevity, the momentum index $k$ was omitted.

The interpolating functions $\mathcal{F}^q_{i + 1/2}$ ($q = 1,2,3$) are given by:
\begin{align}
\mathcal{F}^1_{i+1/2} =& \frac{p_{k}}{m} \left(\frac{1}{3}f_{i-2} - \frac{7}{6} f_{i-1} + \frac{11}{6} f_i\right), \nonumber \\
\mathcal{F}^2_{i+1/2} =& \frac{p_{k}}{m} \left(-\frac{1}{6}f_{i-1} + \frac{5}{6} f_{i} + \frac{1}{3} f_{i+1}\right), \nonumber \\
\mathcal{F}^3_{i+1/2} =& \frac{p_{k}}{m} \left(\frac{1}{3}f_{i} + \frac{5}{6} f_{i+1} - \frac{1}{6} f_{i+2}\right).
\end{align}

The weighting factors $\overline{\omega}_q$ appearing in Eq.~\eqref{eq:weno5_flux_x} are given by:
\begin{equation}
\overline{\omega}_q = \frac{\widetilde{\omega}_q}{\widetilde{\omega}_1+\widetilde{\omega}_2+\widetilde{\omega}_3}, \qquad
\widetilde{\omega}_q = \frac{\delta_q}{\varphi^2_q}.
\label{eq:weno5_omega_x}
\end{equation}

The ideal weights $\delta_q$ are:
\begin{equation}
 \delta_1 = \frac{1}{10}, \qquad \delta_2 = \frac{6}{10},\qquad \delta_3 = \frac{3}{10},
 \label{eq:weno5_delta}
\end{equation}
while the indicators of smoothness $\varphi_q$ can be computed as follows:
\begin{align}
\varphi_1 =& \frac{13}{12} \left(f_{i-2} -2f_{i-1} + f_i \right)^2
+ \frac{1}{4} \left( f_{i-2} - 4f_{i-1} + 3f_i \right)^2,
\nonumber \\
\varphi_2 =& \frac{13}{12} \left(f_{i-1} -2f_{i} + f_{i+1} \right)^2
+ \frac{1}{4} \left( f_{i-1} - f_{i+1} \right)^2,
\nonumber \\
\varphi_3 =& \frac{13}{12} \left(f_{i} -2f_{i+1} + f_{i+2} \right)^2
+ \frac{1}{4} \left( 3f_{i} -4 f_{i+1} + f_{i+2} \right)^2.
\label{eq:weno5_sigma}
\end{align}

\begin{table}
\begin{center}
\begin{tabular}{r|rrr}
 & $\overline{\omega}_1$ & $\overline{\omega}_2$ & $\overline{\omega}_3$ \\\hline
$\phi_1 = \varphi_2 = \varphi_3 = 0$ & $0.1$ & $0.6$ & $0.3$ \\\hline
$\varphi_2 = \varphi_3 = 0$ & $0$ & $2/3$ & $1/3$ \\
$\varphi_3 = \varphi_1 = 0$ & $1/4$ & $0$ & $3/4$ \\
$\varphi_1 = \varphi_2 = 0$ & $1/7$ & $6/7$ & $0$ \\\hline
$\varphi_1 = 0$ & $1$ & $0$ & $0$\\
$\varphi_2 = 0$ & $0$ & $1$ & $0$\\
$\varphi_3 = 0$ & $0$ & $0$ & $1$
\end{tabular}
\end{center}
\caption{The values of the weighting factors $\overline{\omega}_q$ \eqref{eq:weno5_omega_x} when one, two or all three of the indicators of smoothness $\sigma_q$ ($q = 1,2,3$) have vanishing values.\label{tab:weno_x}}
\end{table}

The computation of the weighting factors $\overline{\omega}_q$ \eqref{eq:weno5_omega_x} implies the division between the ideal weights $\delta_q$ \eqref{eq:weno5_delta} and the indicators of smoothness $\varphi_q$ \eqref{eq:weno5_sigma}.
To avoid division by $0$ when either one, two or all three of the indicators of smoothness vanish, we follow Refs.~\cite{BA18,BA19} and compute the weighting factors $\overline{\omega}_q$ directly using Table~\ref{tab:weno_x} in the limiting cases when any of the indicators of smoothness vanishes.

\subsection{\label{sec:gardients} Gradient central difference}

For evaluating the gradients appearing in Eq.~\eqref{eq:j1_expanded} we employ the $6$th order central difference scheme\cite{F88}:
\begin{multline}
 \partial_x Q(x)=\\
 \frac{1}{\Delta x}\left[-\frac{1}{60} Q(x-3\Delta x)+\frac{3}{20} Q(x-2\Delta x)-\frac{3}{4} Q(x-\Delta x)\right.\\
 \left.+\frac{3}{4} Q(x+\Delta x)-\frac{3}{20} Q(x+2\Delta x)+\frac{1}{60} Q(x+3\Delta x)\right]
\end{multline}
where $Q\in\{\ln \rho,u,\ln T \}$.

\section*{References}
\bibliography{bibliography}

\end{document}